\pdfoutput=1 
\documentclass{JINST}

\usepackage{lineno}
\linenumbers

\title{Planar Pixel Sensors for the ATLAS Upgrade: Beam Tests results}

\author{J. Weingarten$^1$\thanks{Corresponding author.}, S. Altenheiner$^2$, M. Beimforde$^3$, M. Benoit$^4$, M. Bomben$^5$, G. Calderini$^{5,6}$, C. Gallrapp$^4$, M. George$^1$, S. Gibson$^4$, S. Grinstein$^7$, Z. Janoska$^8$, J. Jentzsch$^4$, O. Jinnouchi$^9$, T. Kishida$^9$, A. La Rosa$^{10}$, V. Libov$^{11}$, A. Macchiolo$^3$, G. Marchiori$^5$, D. Muenstermann$^4$, R. Nagai$^9$, G. Piacquadio$^4$, B. Ristic$^2$, I. Rubinskiy$^{11}$, A. Rummler$^2$, Y. Takubo$^{12}$, G. Troska$^2$, S. Tsiskaridtze$^7$, I. Tsurin$^{13}$, Y. Unno$^{12}$, P. Weigell$^3$, T. Wittig$^2$
\\
\llap{$^1$}II. Physikalisches Institut, Georg-August-Universit\"at G\"ottingen, Germany \\ Email: \email{jens.weingarten@uni-goettingen.de}\\
\llap{$^2$}Technische Universit\"at Dortmund, Fakult\"at Physik, Dortmund, Germany\\
\llap{$^3$}Max-Planck-Institut f\"ur Physik, M\"unchen, Germany\\
\llap{$^4$}CERN, Gen\`eve, Switzerland\\
\llap{$^5$}Laboratoire de Physique Nucleaire et de Hautes \'Energies (LPNHE), Paris, France\\
\llap{$^6$}Dipartimento di Fisica E. Fermi, Universit\`a di Pisa, and INFN Sez. di Pisa, Pisa, Italy\\
\llap{$^7$}ICREA and Institut de Fisica d'Altes Energies (IFAE), Bellaterra (Barcelona), Spain\\
\llap{$^8$}Division of elementary particle physics - Institute of Physics of the Academy of Sciences of the Czech Republic, Prague, Czech Republic\\
\llap{$^9$}Tokyo Institute Of Technology, Tokyo, Japan \\
\llap{$^{10}$}Section de Physique (DPNC), Universit\'e de Gen\`eve, Gen\`eve, Switzerland\\
\llap{$^{11}$}Deutsches Elektronen-Synchrotron (DESY), Hamburg, Germany\\
\llap{$^{12}$}High Energy Accelerator Research Organization, Tsukuba, Japan\\
\llap{$^{13}$}The University of Liverpool, Liverpool, United Kingdom \\}

\abstract{The performance of planar silicon pixel sensors, in development for the ATLAS Insertable B-Layer and High
Luminosity LHC (HL-LHC) upgrades, has been examined in a series of beam tests at the CERN SPS 
facilities since 2009. Salient results are reported on the key parameters, including the spatial
resolution, the charge collection and the charge sharing between adjacent cells,
 for different bulk materials and sensor geometries.

Measurements are presented for n$^{+}$-in-n pixel sensors irradiated with a range of fluences and
for p-type silicon sensors with various layouts from different vendors. All tested sensors
were connected via bump-bonding to the ATLAS Pixel read-out chip.
      
The tests reveal that both n-type and p-type planar sensors are able to collect significant charge
even after the lifetime fluence expected at the HL-LHC.}

\input{newcommands}


\begin{document}


\section{Introduction}
\label{sec:intro}
The ATLAS collaboration will upgrade the current Pixel Detector~\cite{pixel-electronics-paper} in two phases. 
A first upgrade will be realized during the shut-down in 2013, by inserting a fourth detection layer (Insertable B-Layer - IBL) at a radius of 3.2\,cm from the beam line. The IBL will improve the tracking and vertexing performance of the current pixel detector significantly during operation of the LHC at its nominal centre-of-mass energy ($\sqrt{s}$ = 14\,TeV)~\cite{IBL}.
 
The close proximity to the interaction point imposes a very harsh radiation environment on the IBL. At the end of Phase-I operation of the LHC, foreseen around 2020, the IBL must sustain an estimated fluence of $5\times10^{15}$\,n$_{\rm eq}$/cm$^2$, including a 60\% safety factor at r~=~3.2~cm.

The Phase-II luminosity upgrade for the LHC (beyond 2020) aims to increase the instantaneous luminosity to $5\times10^{34}$\,cm$^{-2}$\,s$^{-1}$, posing a serious challenge to the technology for the ATLAS tracker in the High Luminosity era (HL-LHC): the lifetime fluence for the innermost layer, including safety factors, is estimated to be on the order of $2\times10^{16}$\,n$_{\rm eq}$/cm$^2$~\cite{HL-LHC}. Hence, in view of a possible pixel system replacement after 2020, new pixel sensors are under study.   

Within the Planar Pixel Sensor collaboration (PPS)~\cite{PPS} several optimizations of this well-known technology are under investigation, to address issues arising from the LHC upgrades.
The PPS collaboration investigates the suitability of different materials (p- and n-bulk, diffusion oxygenated float zone, magnetic Czochralski), different geometries (slim edge design, number and width of guard rings) and different biasing / isolation choices (punch-through, polysilicon resistance / p-spray, p-stop), for a new generation of planar pixel sensors. Data taken during beam tests complement tests under laboratory conditions in assessing the performance of various sensor prototypes. In this paper results from two beam test campaigns in 2010 are presented.


The paper is organized as follows. After a description of the experimental setup of the beam tests in Section~\ref{sec:tele} and some details of the data analysis in Section~\ref{sec:analysis}, beam test results are presented on three areas of the scientific program of the PPS collaboration:

Sensors implemented in p-type silicon are being studied by several groups within the PPS collaboration. Section~\ref{sec:n-in-p} shows results from the first beam test operation of various prototype sensors. We will show that the performance of p-type sensors in terms of collected charge~\footnote{collected charge is always presented as Most Probable Value (MPV), if not stated otherwise}, charge sharing probability, and spatial resolution is very similar to that of n-type sensors.

In Section~\ref{sec:n-in-n} the radiation-hardness of sensors implemented in n-type diffusion oxygenated float-zone silicon are studied. We will show that n-in-n detectors are operable at the nominal bias voltage of 1000\,V after a fluence comparable with that expected for IBL ($5\times10^{15}$\,n$_{\rm eq}$/cm$^2$). Noise occupancy, charge collection performance, charge sharing probability and spatial resolution will be presented. 

The new pixel sensors will not only have to sustain the harsher environment, but also have to show high geometrical acceptance without overlapping adjacent modules. Hence the inactive areas of the future pixel sensor have to be reduced significantly. For this reason, efforts were devoted to design detectors with reduced dead area. The ``slim edge'' detector-concept will be presented in Section~\ref{sec:slim}, together with its performance in terms of charge collection at the detectors' edge.



\section{Beam test setup}
\label{sec:tele}
Beam tests are crucial for characterizing the performance of any particle detector. Planar silicon sensors for the ATLAS upgrade have been evaluated in several beam tests in 2009 and 2010. 
Data presented in this paper were taken in two different periods in 2010 at the CERN SPS beamline H6. In both periods pion beams of 120\,GeV/c were used. The high momentum of the beam 
particles minimizes the influence of multiple scattering, enabling high precision tracking using the EUDET beam telescope \cite{EUDET}.

\subsection{The EUDET telescope}
The telescope consists of six equal planes, divided into two groups (arms) of three planes each. The Devices Under Test (DUTs) are mounted in between these two arms of the telescope as well as downstream of the last telescope plane. The increase of the track extrapolation error downstream of the telescope was minimized by mounting the DUTs as close as possible to the last telescope plane.\\
The sensitive elements of the telescope planes are Mimosa26~\cite{Mimosa26} active pixel sensors with a pixel pitch of 18.4\,$\mu$m. Each plane consists of 1152\,$\times$\,576 pixels covering an active area of 21.2$\times$10.6\,mm$^2$. The tracking resolution between the telescope arms is estimated to be 2\,$\mu$m, at the position of the samples downstream of the telescope it is approximately 10\,$\mu$m.~\cite{JBehr} \\
A coincidence of four scintillators (two upstream and two downstream of the telescope) was used for triggering, which resulted in an effective sensitive area of 2$\times$1\,cm$^2$. 

\subsection{Devices Under Test}
All DUTs were read out using the current ATLAS Pixel readout chip (\protect{FE-I3})~\cite{FEI3}. 
The FE-I3 chip is an array of 160 rows $\times$ 18 columns of 50\,$\mu$m $\times$ 400\,$\mu$m read-out cells. In each readout cell the sensor charge signal is amplified and compared to a programmable threshold by a discriminator. The information on the collected signal is encoded through a digital time over threshold (ToT)~\cite{FEI3} measured in units of 25\,ns, which is the nominal LHC bunch crossing rate. The ToT to charge conversion was tuned for each individual pixel to 60\,ToT for a deposited charge of 20\,ke. Discriminator thresholds were tuned to a charge of 3.2\,ke. Prior to the beam test, tuning was performed for every readout chip in realistic conditions, designed to closely resemble those at the beam test. Since the ToT-tuning is particularly temperature dependent, the ToT was calibrated on each sample after installation in the beam test setup to ensure a proper charge conversion. 

For each DUT a fiducial region was defined, based on geometrical and operability considerations. For most studies only the performance of central pixels is of interest. Therefore in many analyses the pixels at the edges of the sensors were masked. In addition, all pixels that were found to have disconnected or merged bump-bonds in laboratory measurements were masked, as were pixels with high noise occupancy.

Devices were irradiated to different fluences using 25~MeV energy protons at the Irradiation Center in Karlsruhe~\cite{KIT}, 24~${\rm GeV/c}$ momentum protons at the CERN PS irradiation facility~\cite{IRRAD1}~\footnote{we observed FE-I3 chip stopped working after being CERN PS irradiated}, and reactor neutrons at the TRIGA reactor of the Jo\v{z}ef Stefan Institute, Ljubljana~\cite{Ljubirrad}. The radiation damage from the different irradiations is scaled to the equivalent damage from 1\,MeV neutrons using the NIEL hypothesis~\cite{NIEL}. During the irradiations the devices were neither powered nor cooled. No standardized annealing procedure was used, but samples were stored below 0~$^{\circ}$C to avoid uncontrolled annealing. The \protect{FE-I3} was designed for a lifetime irradiation dose of 1\,kGy. Some of the DUTs were irradiated to significantly higher doses, leading to increasing numbers of non-working pixels.

For the beam tests, irradiated DUTs were cooled via a strip of copper tape thermally connecting the backside of the DUTs with the base plate of the thermal enclosure. The base plate was in turn cooled using dry ice (CO$_2$)~\cite{Troska}. Due to this setup, the temperature of the sensors varied over time as the dry ice evaporated and needed to be closely monitored. Temperatures were recorded, ranging between -45~$^{\circ}$C shortly after filling the coldbox with dry ice and -15~$^{\circ}$C towards the end of a data taking period.


\section{The beam test analysis}
\label{sec:analysis}

\subsection{Track reconstruction}
The tracks of particles traversing the EUDET telescope are reconstructed from raw hit positions by a sequential algorithm. In the first step, the hits recorded in all telescope planes and DUTs are converted into the EUTelescope~\cite{Eutelescope} internal data format. A time stamp issued by the Trigger Logic Unit~(TLU)~\cite{TLU} is attached to each hit, enabling recovery from any loss of synchronisation between telescope and DUTs during this step, provided the desynchronisation is not too severe. 


As the Mimosa26 sensors of the telescope plane use the rolling shutter read-out technique, the telescope integrates hits for $112\,\mu {\rm s}$ after the arrival of the trigger signal. This is much longer than the 400\,ns hit-buffer of the DUTs, so some tracks will be recorded by the telescope, but not by the DUTs. To correct for this effect, only hits that were recorded within the sensitive time of the DUTs are retained for further analysis. This is done by requesting hits spatially associated to the track in one or more of the other DUTs (in-time tracks). A clustering algorithm is then executed searching for clusters in all planes.

Hits are then transformed from the local coordinate system of each plane to a global coordinate system, where the z-axis gives the beam direction. During this coordinate transformation the pixel sizes in x- and y-directions, the specified z-position of all sensors, and the rotations of the DUTs about all axes are taken into account. Based on correlations between hit positions in different planes in the global frame, a coarse pre-alignment is calculated. Using this information, the alignment processor (see also~\ref{subsec:alignment}) tries to fit tracks through all planes in the setup, taking into account the different spatial resolutions of the planes and individual track selection criteria for each plane. Individual selection criteria are especially necessary, since the pixel size of the telescope-planes is $18.4\,\mu {\rm m}$ in both directions, whereas the investigated FE-I3 based samples have a pixel size of $400\,\mu {\rm m}$ x $50\,\mu {\rm m}$.

The final step is the track-fitting, which is based on a Kalman filter~\cite{DAF}. The track fits are unbiased, requiring a hit in at least four out of the six telescope planes and in at least one DUT. Also in this step, different track selection criteria can be applied. The parameters of all reconstructed tracks are finally stored in a ROOT file~\cite{ROOT} for further analysis (see Section ~\ref{subsec:tbmon}).

\subsection{Detector alignment}
\label{subsec:alignment}

The alignment for the telescope planes and the DUTs in the EUTelescope track-reconstruction uses the MILLIPEDE tool~\cite{Millipede}. In the algorithm the alignment constants are calculated such that the uncertainties of the fitted track parameters, as well as the $\chi^2$ of the track residuals, are minimized. Straight line fits to the hit positions in all active planes are performed independently for the x- and y-directions. Individual criteria can be applied to the resulting residual distributions to suppress fake tracks. In the alignment process the pre-alignment constants, calculated in the previous hitmaker step, are taken into account. This enables alignment of all telescope-planes and DUTs in one step, where just the first telescope-plane is fixed in its position and orientation. The alignment constants are applied in the final track fitting process.

\subsection{Data analysis}
\label{subsec:tbmon}
The analysis of the reconstructed tracks is conducted in several steps, using a dedicated data analysis framework (tbmon)~\cite{tbmon}.

Firstly, unresponsive and noisy pixels are identified and masked. A pixel is unresponsive if it registers no hit during the full data taking period and noisy if more than  $5\times10^{-4}$ of all hits registered in this pixel are not correlated with a beam particle. 
Typically, less than 1\,\% of pixels are masked in non-irradiated modules; for irradiated 
devices the fraction fluctuates between samples. On average, roughly 10\,\% of the pixels have to be masked due to problems with settings of the readout chip or increased noise due to high leakage current of the sensor.



In the following analyses tracks extrapolated from the telescope are ``matched'' to a hit if the hit and the extrapolated track impact point in the DUT plane containing the hit are closer than 400\,$\mu$m  in the long pixel direction and 150\,$\mu$m  in the short pixel direction. The hit position is defined as the $\eta$ corrected ToT weighted position of all pixels in a cluster~\cite{Eta}. 

To estimate the intrinsic spatial resolution of the DUTs the distribution of hit residuals is studied. The hit residual is defined as the distance between the reconstructed hit position on the DUT and the extrapolation of the fitted track to the DUT plane. The intrinsic spatial resolution is estimated by the RMS of the residual distribution for clusters of all sizes, while the residual distribution of 2-pixel clusters is used to estimate the width of the area between pixels, where charge sharing occurs. The distribution is fitted with the sum of two Gaussian functions, where one accounts for mis-reconstructed hits, resulting in large residual values (equal to 2 times the pixel pitch or more), and the other for correctly reconstructed hits. The width of this ``core'' Gaussian gives the width of the charge sharing region.

To calculate the charge sharing probability for each hit within a cluster, it is determined whether a hit is found in a pixel cell adjacent to the one matched to a track. 
This probability increases towards the edge of the pixel since charge carriers are more likely to drift to the neighbouring pixel. The corresponding plot, referred to as a charge sharing map, is centred on one pixel, also showing half of the adjacent pixel in each direction. 
The overall charge sharing is defined as the number of tracks with at least one hit in a neighbouring pixel divided by the number of all tracks.

Due to a problem in the readout system, random DUTs stopped sending data for random short intervals. Therefore, the availability of a reference plane for the selection of in-time tracks cannot be ensured at all times. As this selection is crucial for the measurement of the hit efficiency, this analysis could not be done with the available data, while charge collection and spatial resolution measurements are unaffected.

\begin{figure}[!htb]
\begin{center}
\includegraphics[width=\textwidth]{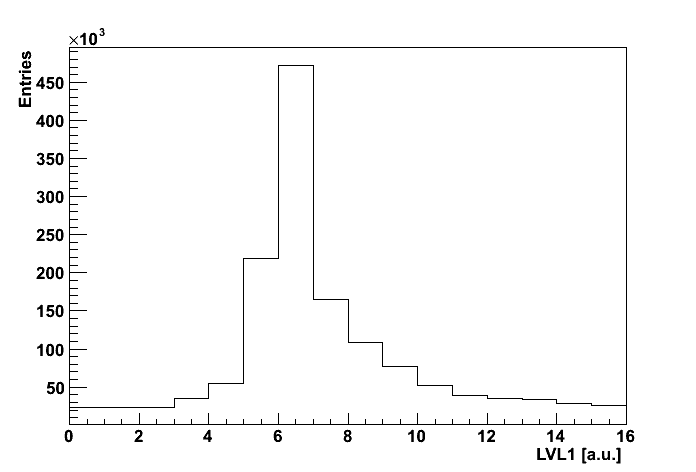}
\caption{\label{fig:lvl1}Example of LVL1 distribution.}
\end{center}
\end{figure}

 Most of the hits registered by the DUTs were anyway associated with tracks; this 
can be seen from the LVL1 distribution.
The LVL1 distribution  (see Figure~\ref{fig:lvl1}) shows the arrival time of every recorded hit with
respect to the external trigger signal. The very pronounced peak shows that
most hits have a strong correlation with the timing of the external trigger
signal, meaning that they are indeed generated by the triggered particle
traversing the DUT. By applying cuts to the LVL1 distribution, we can therefore suppress most hits
that are not associated to a track.

\section{The \ninp\ demonstrator program}
\label{sec:n-in-p}
While \ntype\ bulk sensors require patterned guard rings on the back side of the sensor, for \ptype\ material these can be moved to the pixelated side of the sensor (front side); then metallization is the only process for the back side. This makes it a very cost-effective material for future pixel detectors. On the other hand the high voltage, which is applied to the back side of the sensor, is also present on the edges of the front side of the sensor facing the read-out  chip, which is at ground potential. Thus spark discharges may occur, posing a risk to the readout electronics itself. This can be limited by the deposition of an insulating coating on the edge of the sensor; more details are given in the next section.\\
Sensors in p-type technology tested in 2010 were produced at CiS\footnote{Forschungsinstitut f\"ur Mikrosensorik und Photovoltaik GmbH}  and at HPK\footnote{Hamamatsu Photonics K. K.}. Table \ref{tab:p-samples} summarizes the relevant quantities for the devices studied in the beam test. Further details of each sample are described below.

\begin{table}[!htb]
\begin{center}
\begin{tabular}{l|c|c|c|c}
sample & Period & fluence ($10^{15}$\,n$_{\rm eq}$/cm$^2$) & irradiation type & $V_{bias}$ [V]\\
\hline
\hline
MPP1 &July & 0 & --       & 150, 200                \\
MPP2 & July & 0 & --       & 150, 200                \\
MPP3 & October & 0 & --       & 150                     \\
MPP4 & October &1 & reactor neutrons & 250, 350, 500, 550, 700 \\
MPP5 & October &1 & 25~MeV protons  & 500, 550                \\
KEK1 &October & 0 & --       & 100, 200                \\
KEK2 &October & 0 & --       & 100, 200                
\end{tabular}
\end{center}
\caption{\label{tab:p-samples}Relevant quantities of the n-in-p samples.}
\end{table}

\subsection{CiS sensors}
In the following the p-type sensors produced at CiS will be introduced and their beam test results discussed.

\subsubsection{Sensor design}

The n-in-p pixel sensors labelled as MPP1--MPP5 were produced at CiS with a geometry compatible with FE-I3, in the framework of a common RD50 production~\cite{n-in-p-CiS}.
They were made from Diffusion Oxygenated Float Zone (DOFZ) 285~$\mu$m thick wafer,
 with $<$100$>$ crystal orientation and a wafer resistivity of 15~k$\Omega$cm.
The depletion voltage before irradiation was nominally 60\,V.

\begin{figure}[!htb]
\begin{center}
\includegraphics[width=0.45\textwidth]{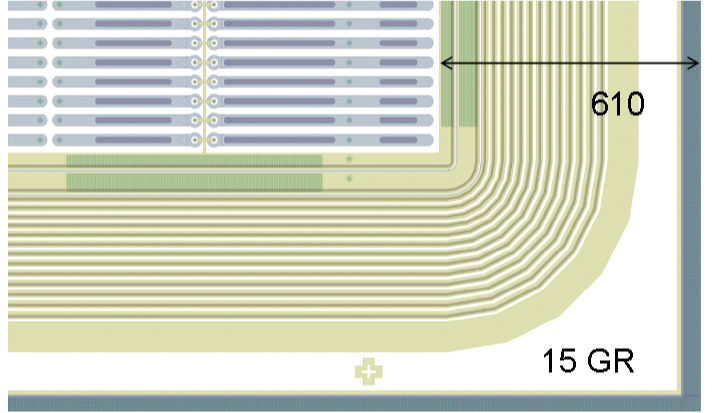}
\includegraphics[width=0.40\textwidth]{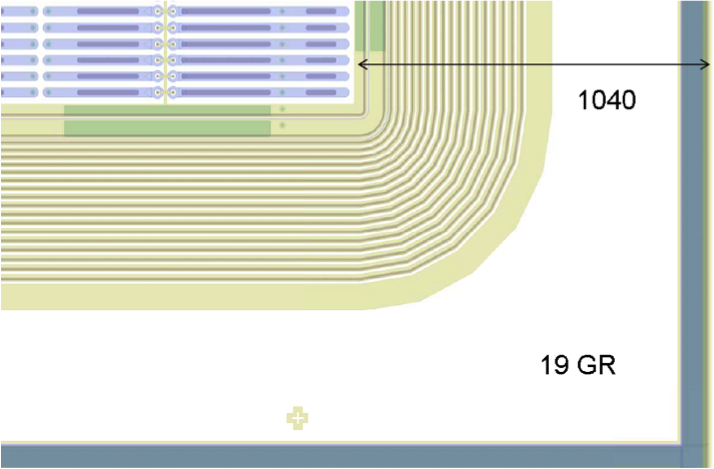}
\caption{\label{fig:n_in_p-CiS}Left: View of a corner of an n-in-p sensor of the CiS production with 15 guard rings. Right: View of a corner of an n-in-p sensor of the CiS production,with 19 guard rings.}
\end{center}
\end{figure}

Two guard rings structures with differing widths have been implemented and tested. One design has the standard inactive area of 1\,mm per side of normal ATLAS pixel sensors, while the other has a reduced inactive area, as illustrated in Figure~\ref{fig:n_in_p-CiS}. MPP1, MPP3 and MPP5 have 19 guard-rings, with the standard total inactive width of 1 mm on each side. The samples MPP2 and MPP4 (see Table~\ref{tab:p-samples}) have 15 guard-rings with a total inactive area of 610\,$\mu$m on each side.
For both guard rings designs the bias ring and the inner guard ring are wider
 than
 the other ones, enabling tests of the sensors before connection to
 the read-out chips.
The external guard ring widths grows with promixity to the cutting edge from
 17~$\mu$m to 22~$\mu$m, with the gap between the rings from 5 to 8~$\mu$m.
The distance between the last ring and the dicing edge is 400~$\mu$m for the
 19 guard rings design while it is 100~$\mu$m in the 15 guard rings design.

 The inter-pixel isolation is achieved by means of a homogenous p-spray implantation.

\begin{figure}[!htb]
\begin{center}
\includegraphics[width=0.45\textwidth]{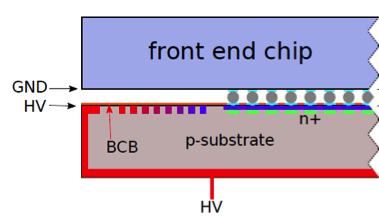}
\caption{\label{fig:n_in_p-BCB}Schematics of a CiS n-in-p pixel assembly. The potential of the different parts is given. The BCB layer is indicated in orange.}
\end{center}
\end{figure}

A BCB (Benzocyclobutene) coating has been applied to the pixelated side of the n-in-p pixel sensors, to prevent sparks between the area outside the guard ring area, that is at the same high 
potential as the back side, and the chip, at ground potential (Figure~\ref{fig:n_in_p-BCB}). The interconnection to the chips has been performed via bump-bonding at IZM-Berlin~\footnote{Fraunhofer-Institut f\"ur Zuverl\"assigkeit und Microintegration, Berlin}.

\subsubsection{Beam test results}
As documented in Table~\ref{tab:p-samples}, during the first beam test period (July 2010) none of the CiS n-in-p sensors (namely MPP1, MPP2) were irradiated. In the second period (October 2010) MPP4 was tested after irradiation with reactor neutrons~\cite{Ljubirrad} to a fluence of  $1\times10^{15}$\,n$_{\rm eq}$/cm$^2$ and  MPP5 was irradiated with low energy protons (25~MeV) at the cyclotron of the Karlsruhe Institute of Technology (KIT)~\cite{KIT} to the same equivalent fluence. MPP3 was kept as an unirradiated reference. The performance of these five samples is presented as follows.

\paragraph{Cluster size}
The cluster size distribution was studied for all sensors as a function of the bias voltage. A detailed breakdown is reported in Table~\ref{tab:p-matchedCSOctober}. For the non-irradiated sample roughly 70\,\% of the clusters consisted of a single hit. A further 25\,\% were two-hit clusters, while the remainder were three or more hit clusters.

\begin{figure}[!htb]
\begin{center}
\includegraphics[width=0.96\textwidth]{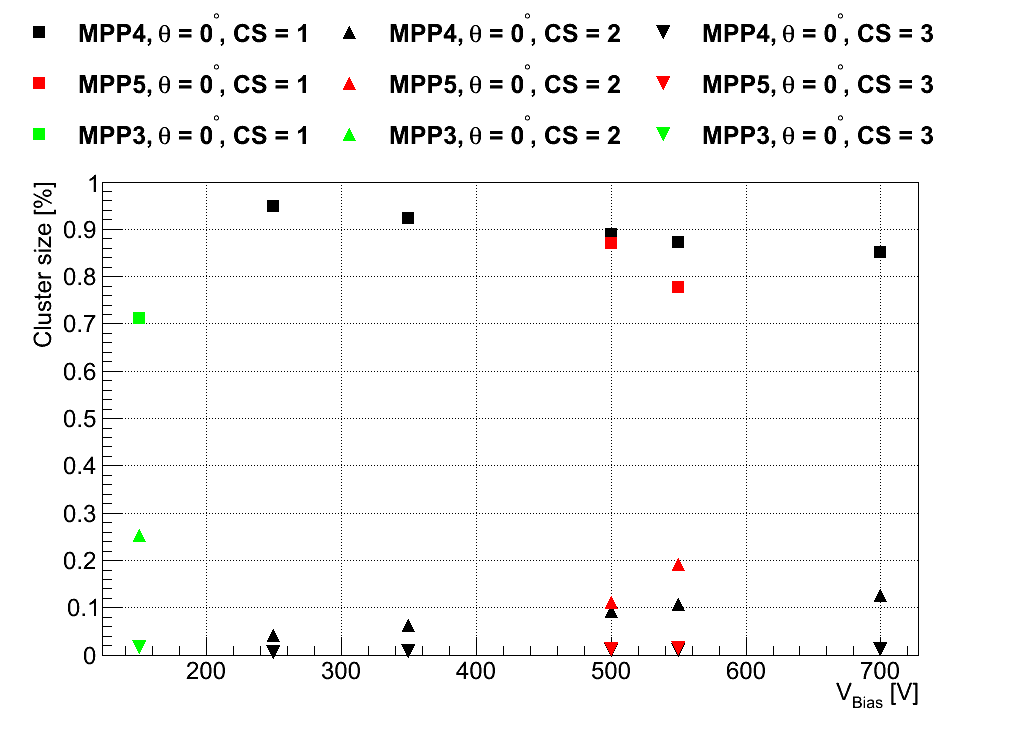}
\end{center}
\caption{\label{fig:mpp_radComp}Relative cluster size abundance as a function of bias voltage for irradiated MPP4 and MPP5 devices. 
MPP3 (non-irradiated) are added for comparison. Particles were impinging at normal incidence. Errors on fractions are negligible and so not visible in the plot.}
\end{figure}

Due to trapping effects and lower overall charges for irradiated sensors, one hit clusters are more often observed than in unirradiated samples. With increasing bias voltage the number of two hit clusters rises since it becomes more likely that a neighbouring pixel is above threshold as well. This behaviour can be clearly seen in Figure~\ref{fig:mpp_radComp}.

\begin{table}[!htb]
\begin{center}
\begin{tabular}{l|c|c|c|c|c}
sample &  $V_{bias}$ [V] & CS=1 [\%] & CS=2 [\%] & CS=3 [\%] & \specialcell[t]{Charge sharing\\ 
probability [\%] }\\
\hline
\hline
MPP1 & 150 & 65& 31 & 2 & 33\\
MPP1 & 200 & 60& 36& 2&  37\\
MPP2 & 200 & 75& 23 & 1&  24\\
MPP3 & 150 & 71 & 25 & 1 & 27\\
\hline
MPP4 & 250  & 95 & 4 & 1&4\\
MPP4 & 350  & 92 & 6 & 1&7\\
MPP4 & 500  & 88 & 9 & 1&10\\
MPP4 & 550  & 87 & 11 & 1 & 11\\
MPP4 & 700  & 85 & 13 & 1&14\\
\hline
MPP5 & 500  & 87  & 11 & 1 & 12\\
MPP5 & 550  & 78 & 19 & 1& 21\\
\hline
\end{tabular}
\end{center}
\caption{\label{tab:p-matchedCSOctober}
Cluster size (CS) composition for CiS modules measured at different bias voltages during beam tests;
  clusters were matched to a track. 
Charge sharing probability is also reported.}
\end{table}

\begin{figure}[!htb]
\begin{center}
\includegraphics[width=1\textwidth]{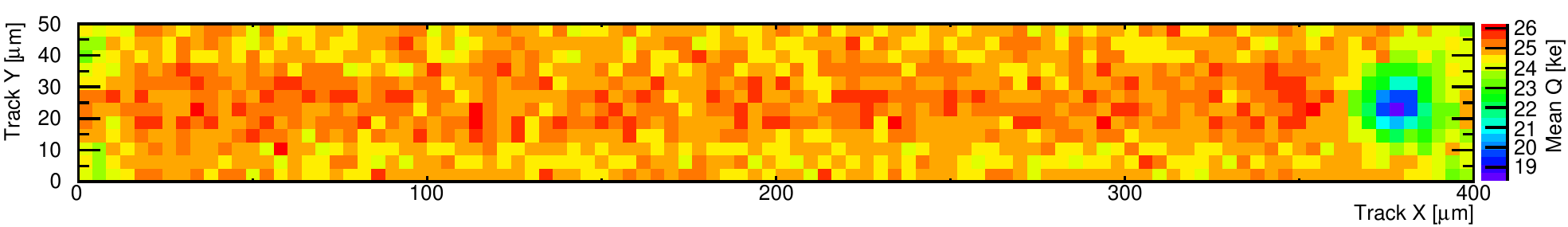}
\end{center}
\caption{\label{fig:chargemapsampleMPP3}Charge collection within a single pixel by track position for MPP3 at $V_{\mathrm{bias}}$ of 150\,V}
\end{figure}

\begin{figure}[!htb]
\begin{center}
\includegraphics[width=1\textwidth]{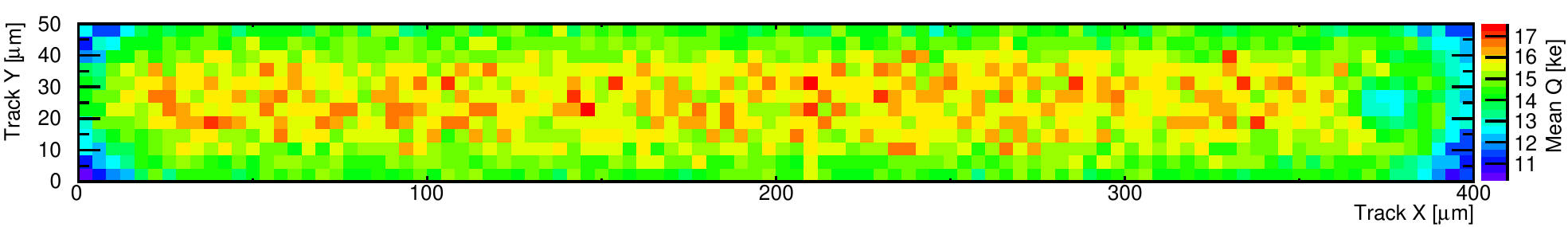}
\end{center}
\caption{\label{fig:chargemapsampleMPP2Oct}Charge collection within a single pixel by track position for MPP5 at $V_{\mathrm{bias}}=500$\,V}
\end{figure}

\paragraph{Collected charge}
The collected charge was measured as a function of the bias voltage for the different sensors.
In Figure\,\ref{fig:chargemapsampleMPP3} the sub-pixel resolved charge collection profile is shown for MPP3 at $V_{\mathrm{bias}}$  of 150\,V. The most prominent feature is the lower collected charge value on the  right hand side, corresponding to the bias dot region where the pixel implant is connected to the bias grid. The same effect is evident in \ninn\ devices with the same design (Section~\ref{sec:n-in-n}). 
In this region the collected charge is still well above the threshold.\\ 
Figure~\ref{fig:chargemapsampleMPP2Oct} shows lower collected charge along the edges of a pixel, due to charge sharing with the neighboring pixels. As charge sharing occurs, less charge is available for the pixel traversed by the particle, decreasing the probability to pass the electronics threshold. This effect is especially pronounced in the corners of the pixel, as charge can be shared among four pixels.
However, in these regions the deposited charge is still high enough for an efficient operation of the device. 

\begin{figure}[!htb]
\begin{center}
\includegraphics[width=0.85\textwidth]{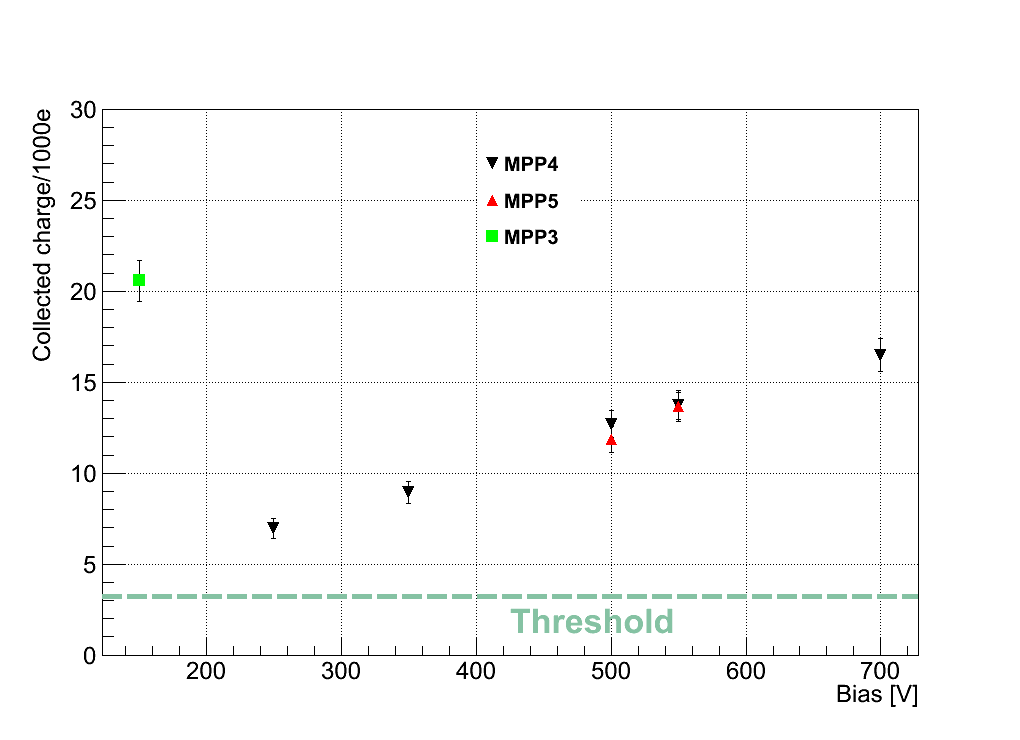}
\end{center}
\caption{\label{fig:MPV_MPP_radComp.png}Charge collected in a cluster: 
most probable value of the charge distribution fitted to a Landau function convoluted with a gaussian as a function of the 
bias voltage. See text for the discussion on the assigned systematic uncertainty. The threshold value is also depicted as a dashed line.}
\end{figure}

Figure~\ref{fig:MPV_MPP_radComp.png} shows the most probable charge for all samples. For comparison the charge collected by unirradiated devices (MPP1 and MPP2) is included and a typical discriminator threshold of 3200\,e is indicated. A systematic error on the collected charge of 400~e is assumed, due to the finite charge resolution of the ToT mechanism; a 5\% systematic uncertainty is taken into account, due to non-uniformity in the injection capacitances. 

Although the irradiated samples do not show saturation of the collected charge up to 700\,V, already at low bias voltages the collected charge exceeds the electronics threshold by more than a factor of two and can thus be considered safe for tracking applications.

\begin{figure}[!htb]
\begin{center}
\includegraphics[width=1\textwidth]{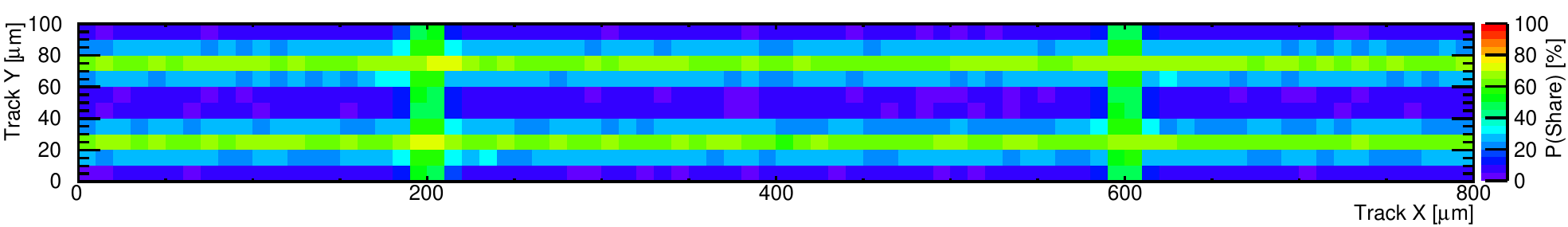}
\includegraphics[width=1\textwidth]{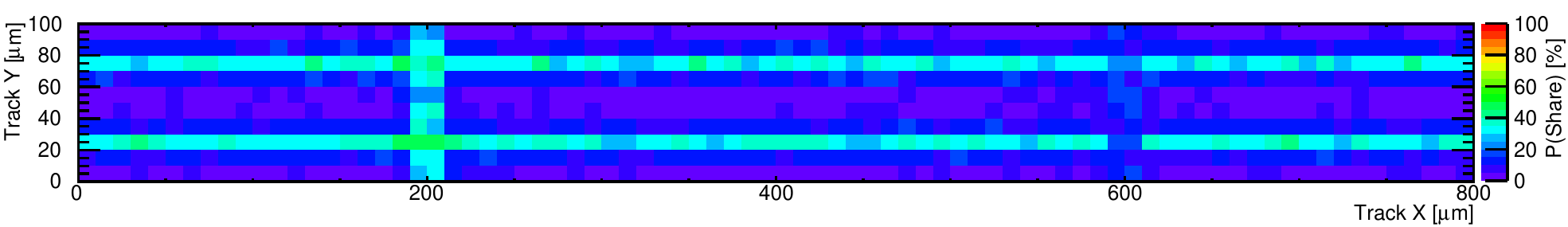}
\end{center}
\caption{\label{fig:MPP3qshare}Charge sharing map for MPP3 detector, biased at 150\,V (top) and for MPP4 detector, biased at 700\,V.} 
\end{figure}

\paragraph{Charge sharing}
In Figure~\ref{fig:MPP3qshare} (top) the charge sharing probability within one pixel for MPP3 is shown. At normal track incidence increased charge collection probability is evident at the edges and corners of the pixel. The situation after irradiation is shown in Figure~\ref{fig:MPP3qshare} (bottom). Here the charge sharing especially on the side of the punch through biasing is reduced, since there is a higher probability for the neighbouring pixel to be below threshold. This is also reflected in the average charge sharing probability given in Table~\ref{tab:p-matchedCSOctober} for the n-in-p CiS detectors in all states. With increasing bias voltage an increase of the charge sharing is observed for the irradiated sensors (MPP4, MPP5) due to the increase in the collected charge. 

\paragraph{Residuals}

\begin{figure}[!htb]
\begin{center}
\includegraphics[width=0.45\textwidth]{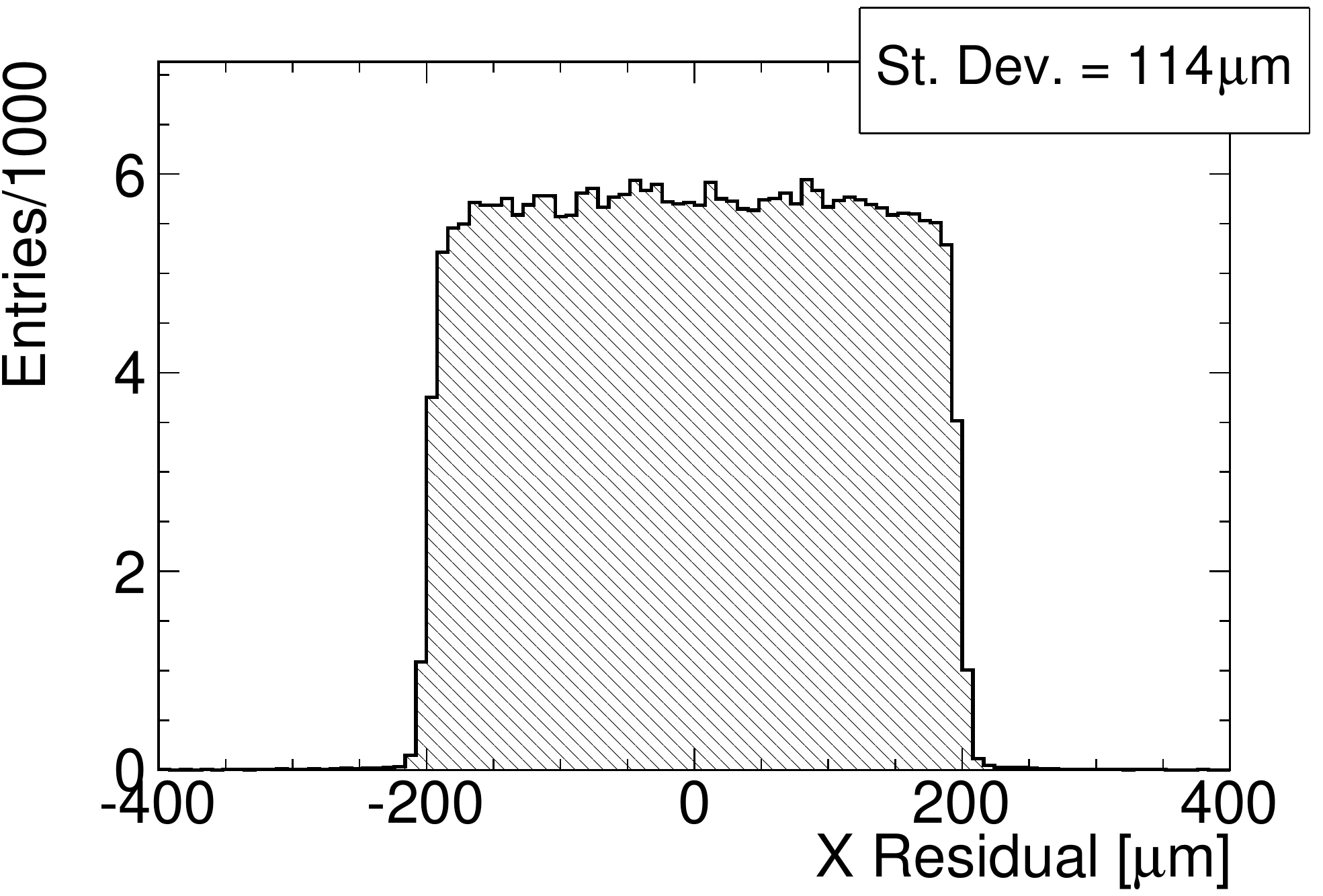}
\includegraphics[width=0.45\textwidth]{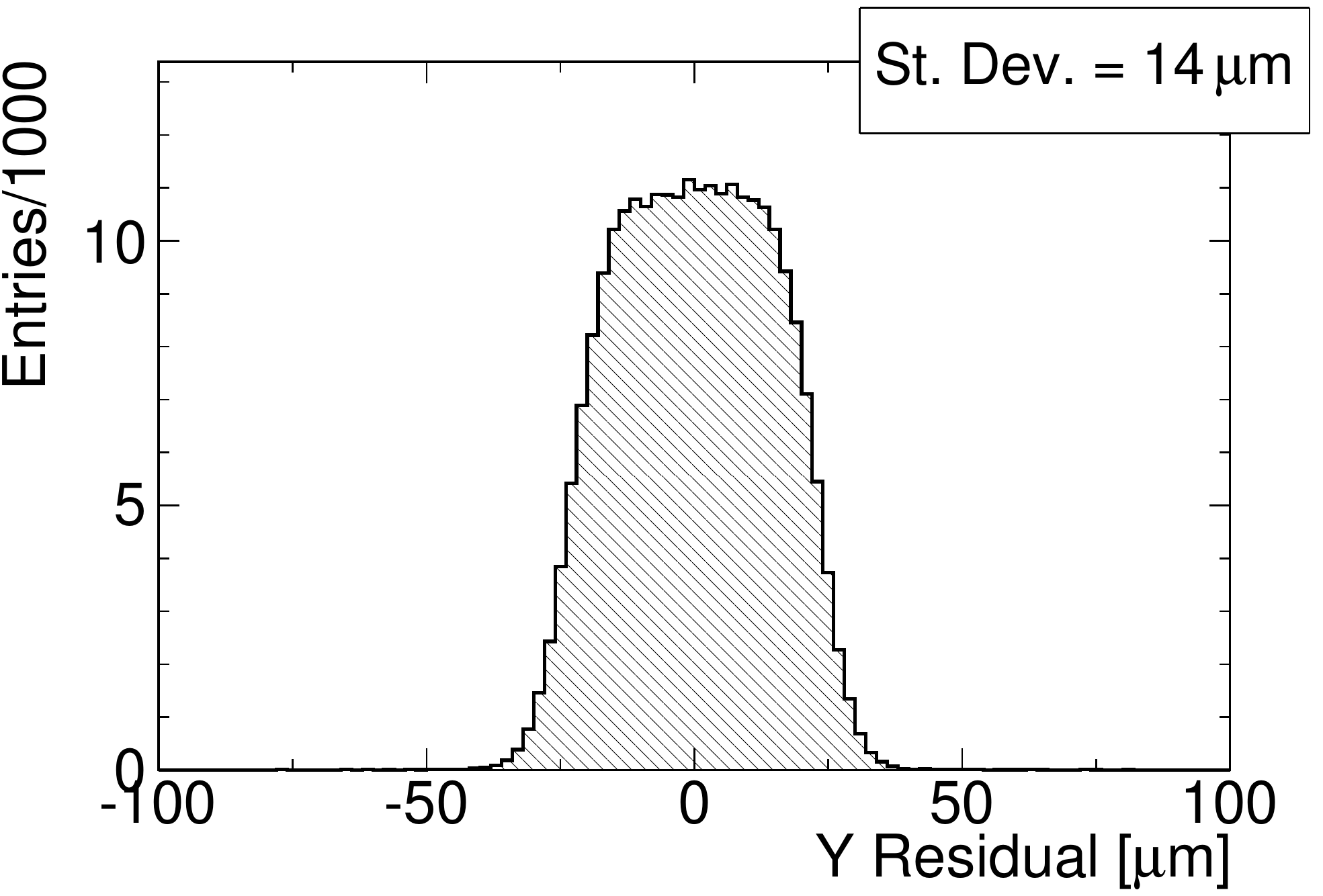}
\end{center}
\caption{\label{fig:MPP5_500_resid}Residual distribution for irradiated sample MPP5 at V$_{bias}$ of 500\,V. Left: long pixel projection; right: short pixel projection.}
\end{figure}

\begin{figure}[!htb]
\begin{center}
\includegraphics[width=0.45\textwidth]{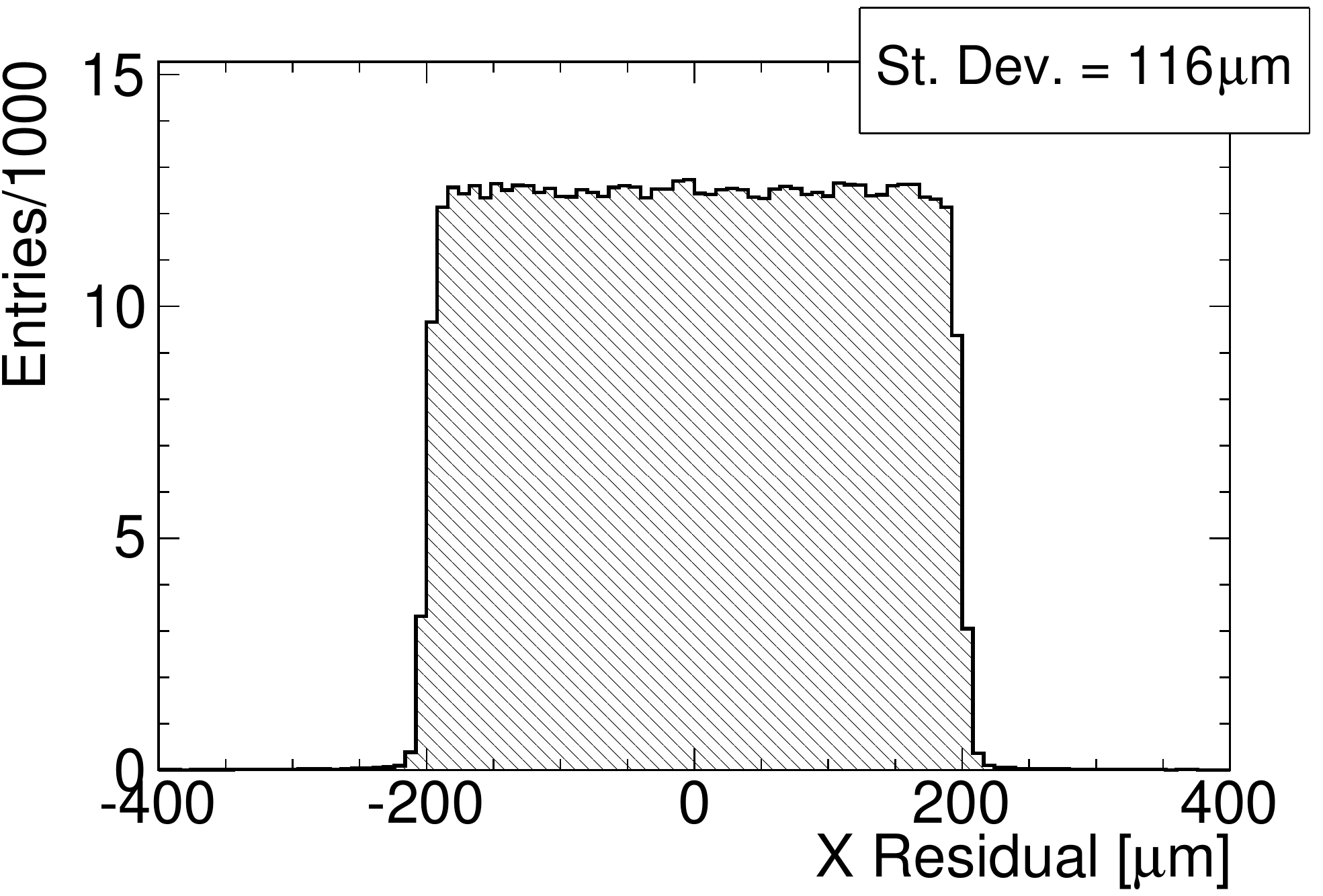} 
\includegraphics[width=0.45\textwidth]{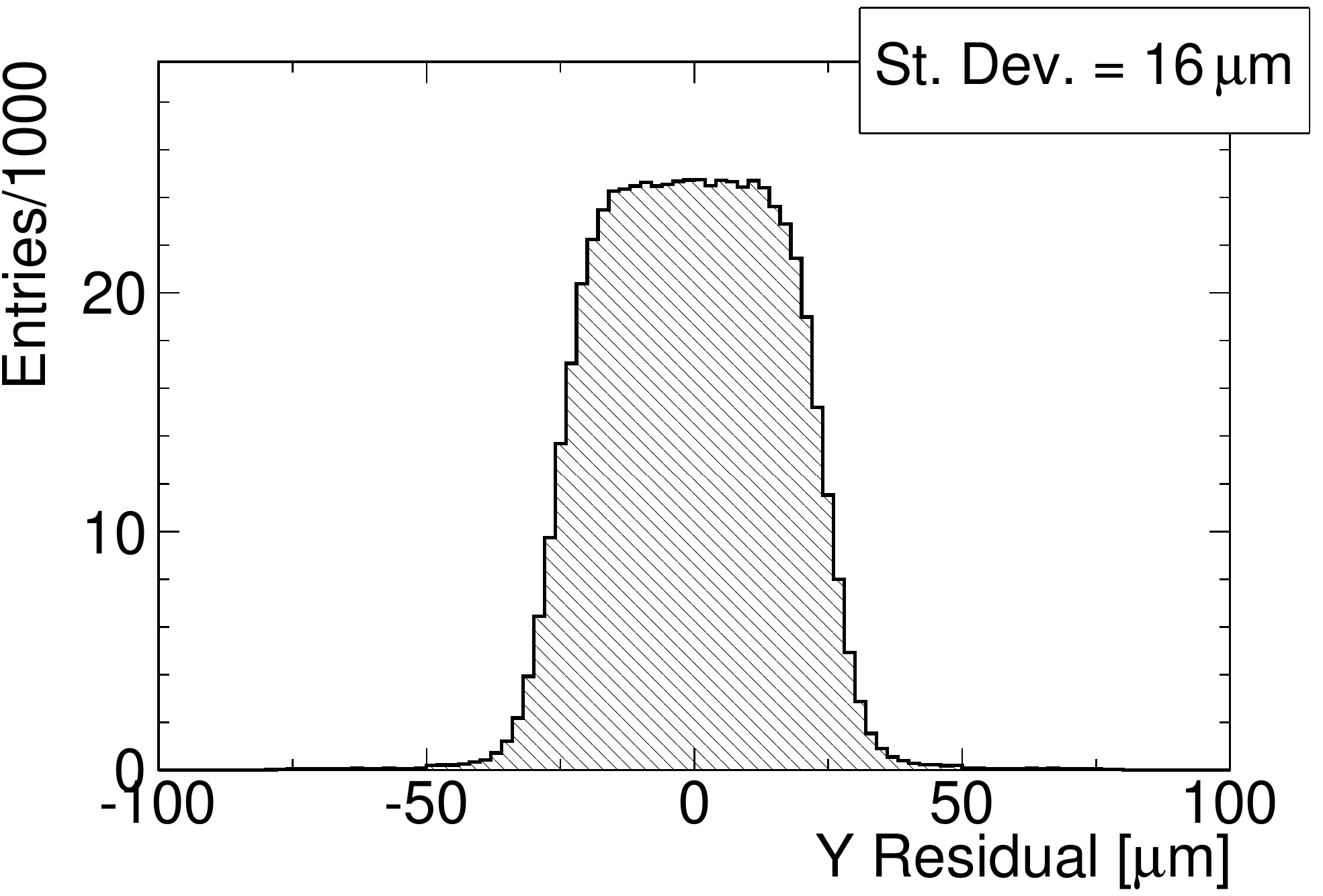} 
\end{center}
\caption{\label{fig:MPP3_150_resid}Residual distribution for non-irradiated MPP3 biased at 150 Volts. Left: long pixel projection; right: short pixel projection.}
\end{figure}

\noindent Figure~\ref{fig:MPP5_500_resid} shows the cluster position residual distribution for the irradiated module MPP5, biased at 500\,V. The spatial resolution is compatible with the digital resolution, as one-hit clusters are dominant. For comparison, the residual distribution for MPP3 at 150\,V is shown in Figure~\ref{fig:MPP3_150_resid}. No difference is appreciable between the two samples.

\begin{figure}[!htb]
\begin{center}
\includegraphics[width=0.45\textwidth]{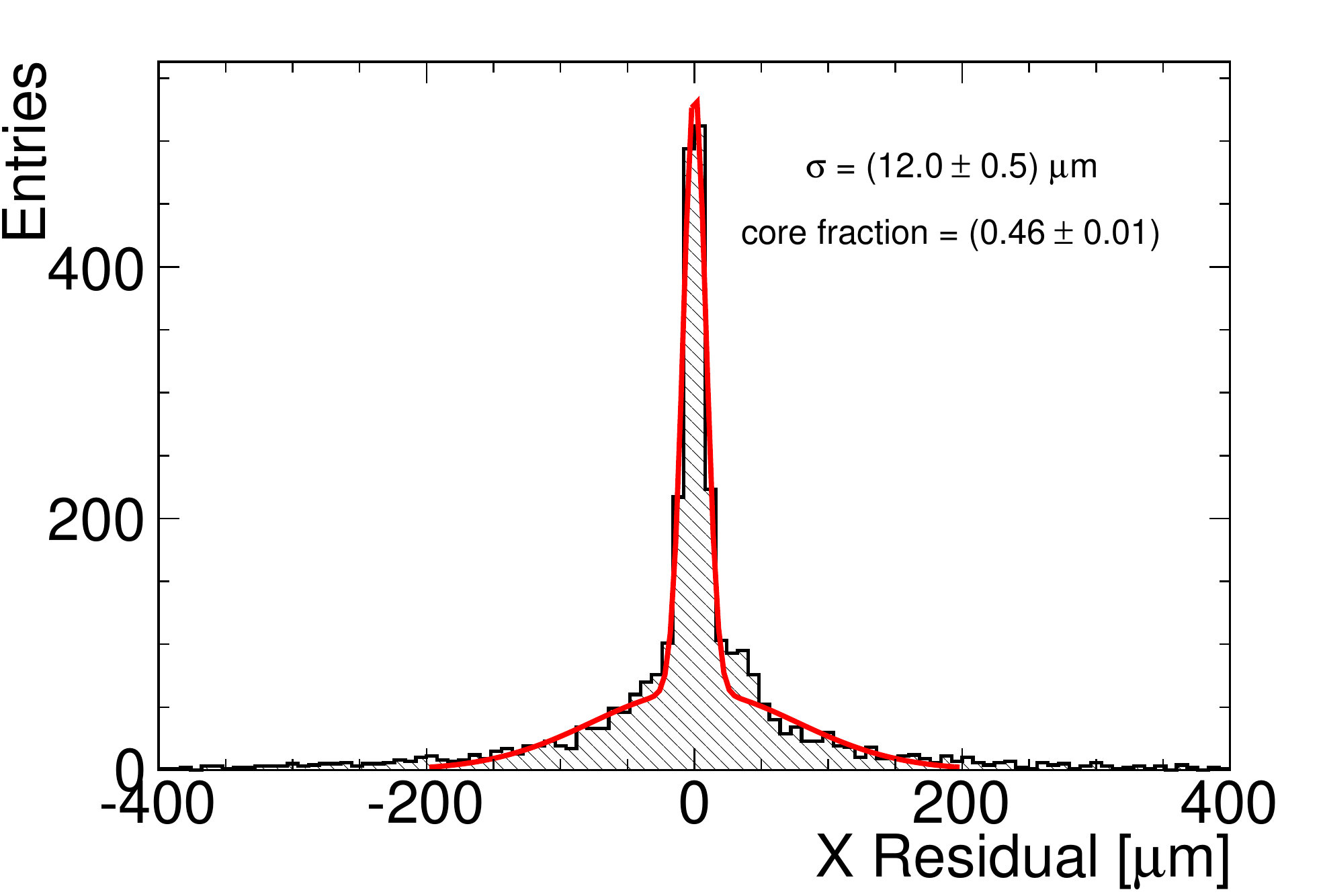} 
\includegraphics[width=0.45\textwidth]{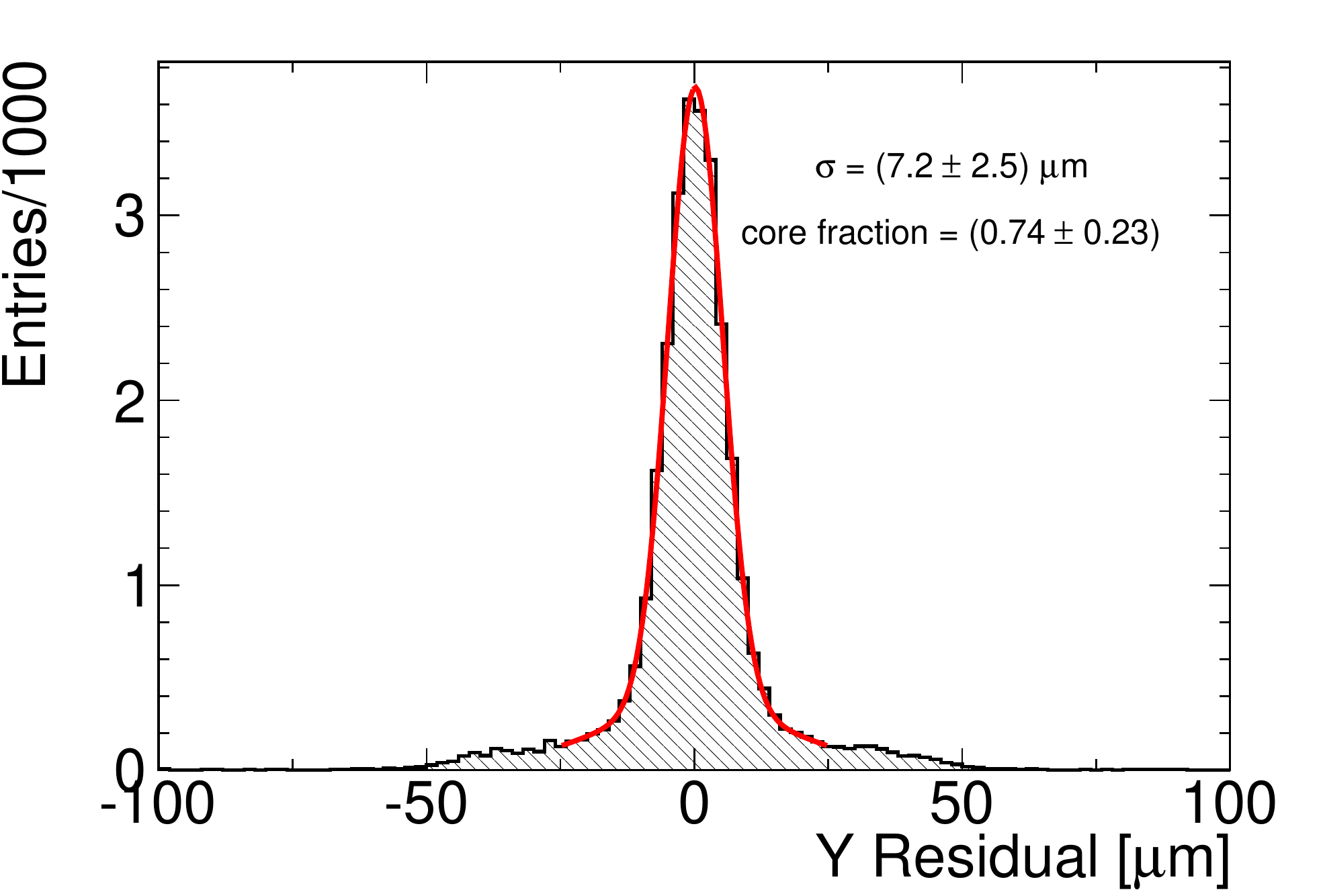} \end{center}
\caption{\label{fig:MPP5_500_resid2}Residual distribution for irradiated sample MPP5 biased at V$_{bias}$ of 500\,V, for two-hit clusters. Left: long pixel projection; right: short pixel projection.}
\end{figure}

\begin{figure}[!htb]
\begin{center}
\includegraphics[width=0.45\textwidth]{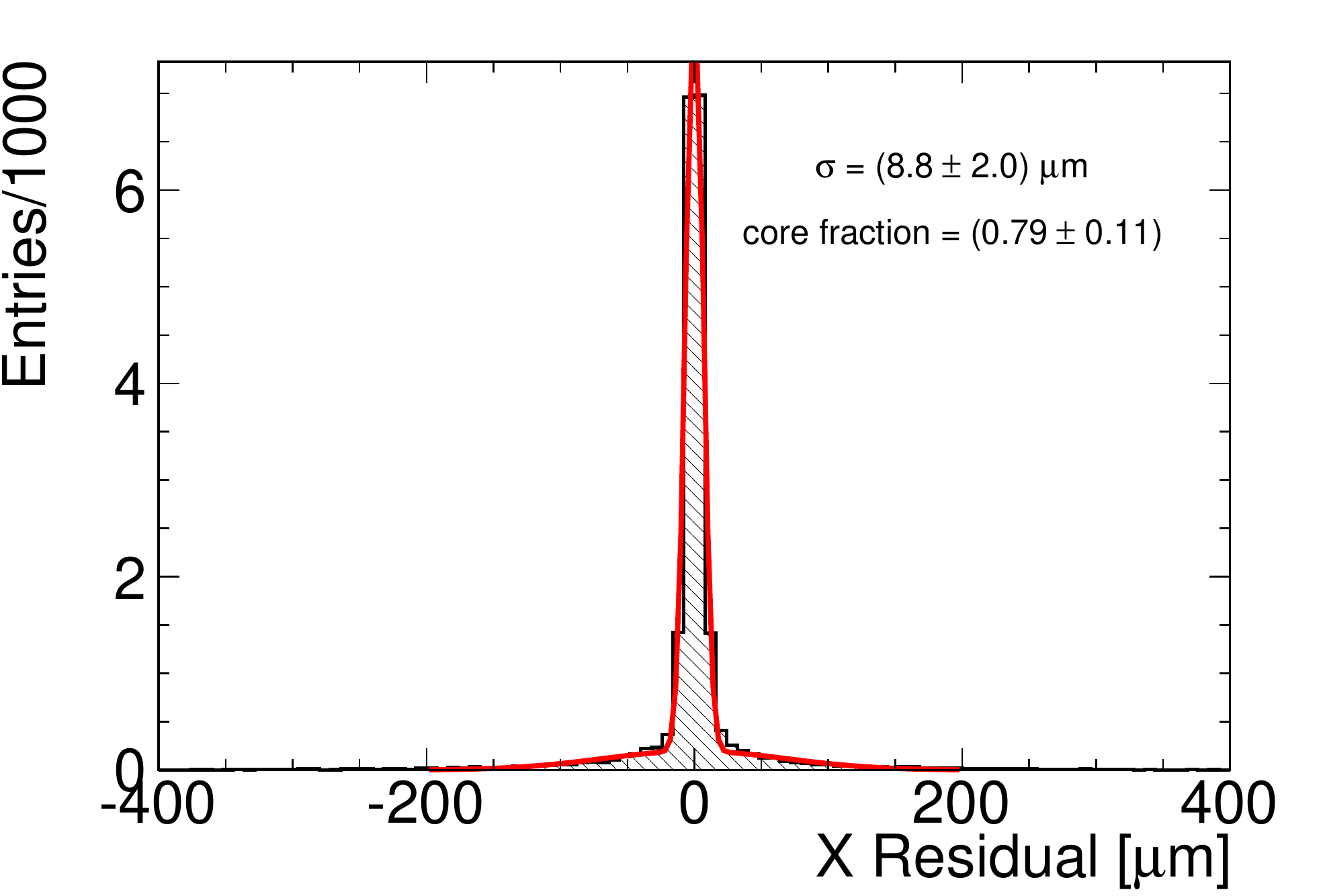} 
\includegraphics[width=0.45\textwidth]{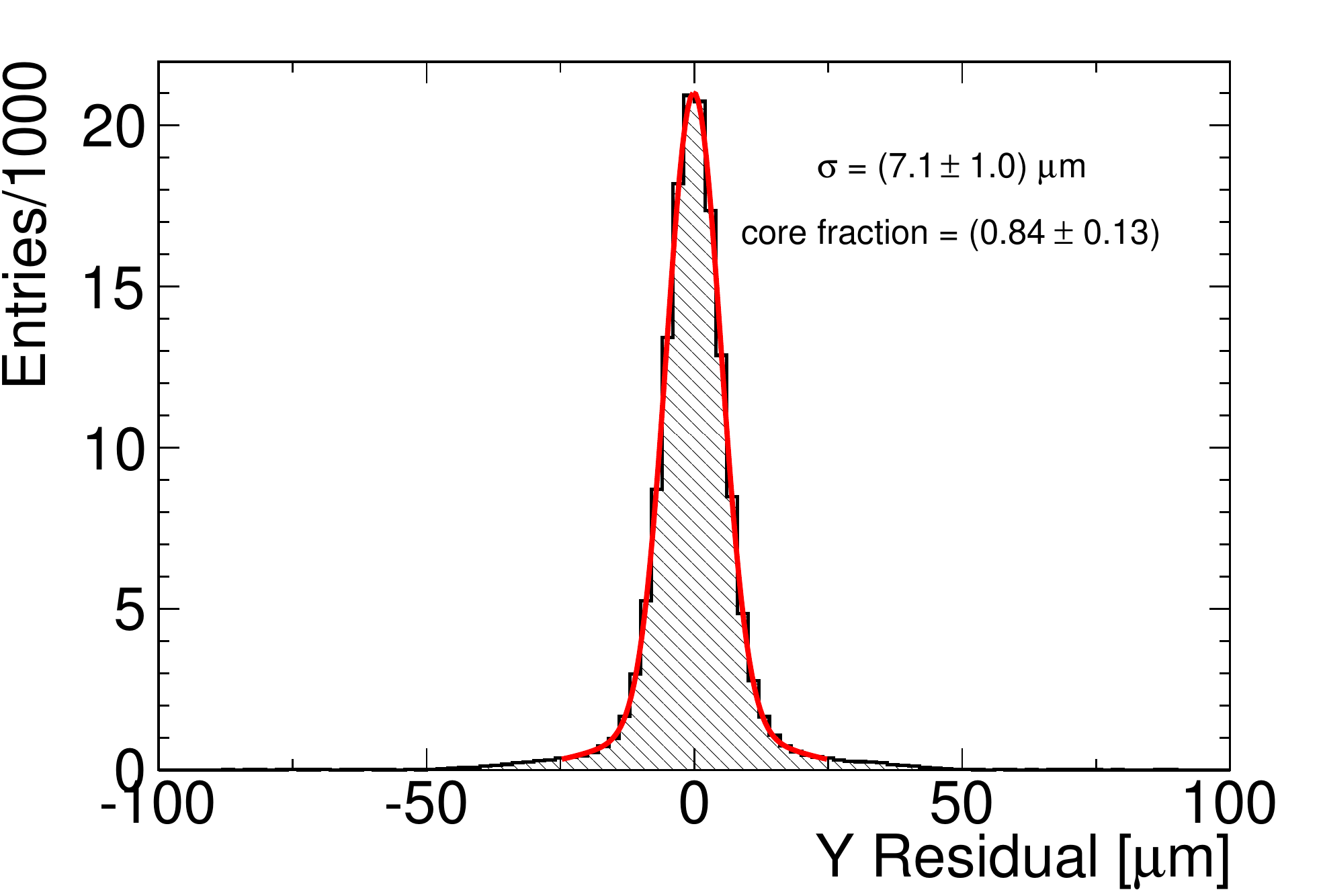}
\end{center}
\caption{\label{fig:MPP3_150_resid2-here}Residual distribution for unirradiated sample MPP3 at V$_{bias}$ of 150\,V, for two-hit clusters. Analysis restricted to clusters with 2 pixels. Left: long pixel projection; right: short pixel projection.}
\end{figure}

For the irradiated assembly MPP5 Figure~\ref{fig:MPP5_500_resid2}  shows in case of two hit clusters a resolution of $(7.2\pm2.5)\mu$m in the core gaussian along the short pixel direction. For comparison, Figure~\ref{fig:MPP3_150_resid2-here} shows the two-hit cluster residuals for MPP3 at 150\,V. After irradiation there are more noise related hits, clearly visible in Figure~\ref{fig:MPP5_500_resid2}. The fitted core fraction indeed decreases
with respect to unirradiated sample (Figure~\ref{fig:MPP3_150_resid2-here}). Nonetheless, the tracking capabilities of irradiated DUTs, in terms of  spatial resolution, are still satisfactory.

\subsection{HPK sensors}
In the following the p-type sensors produced at HPK will be introduced and their beam test results discussed.

\subsubsection{Sensors design}

\begin{figure}[!htb]
\begin{center}
\includegraphics[width=0.65\textwidth]{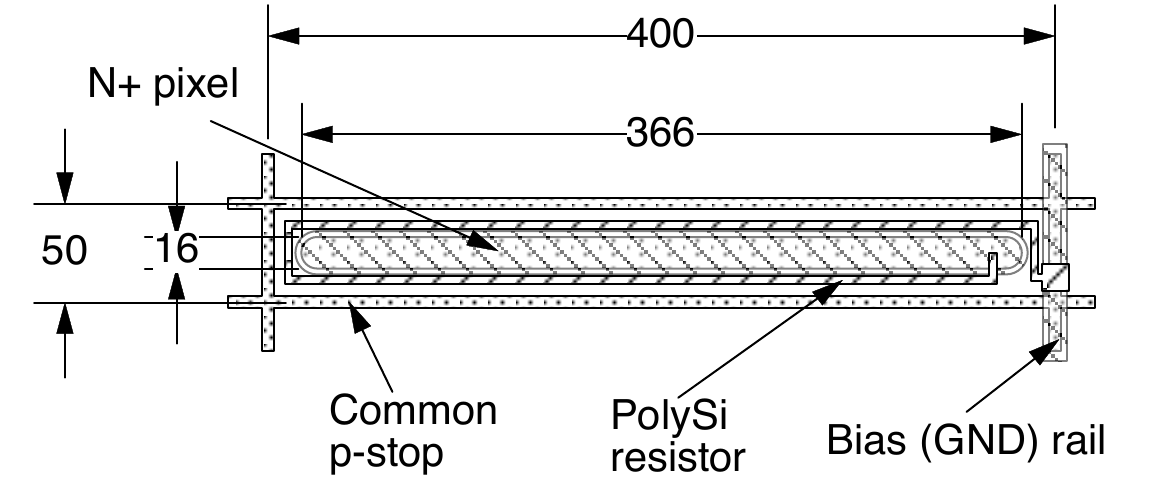}
\includegraphics[width=0.65\textwidth]{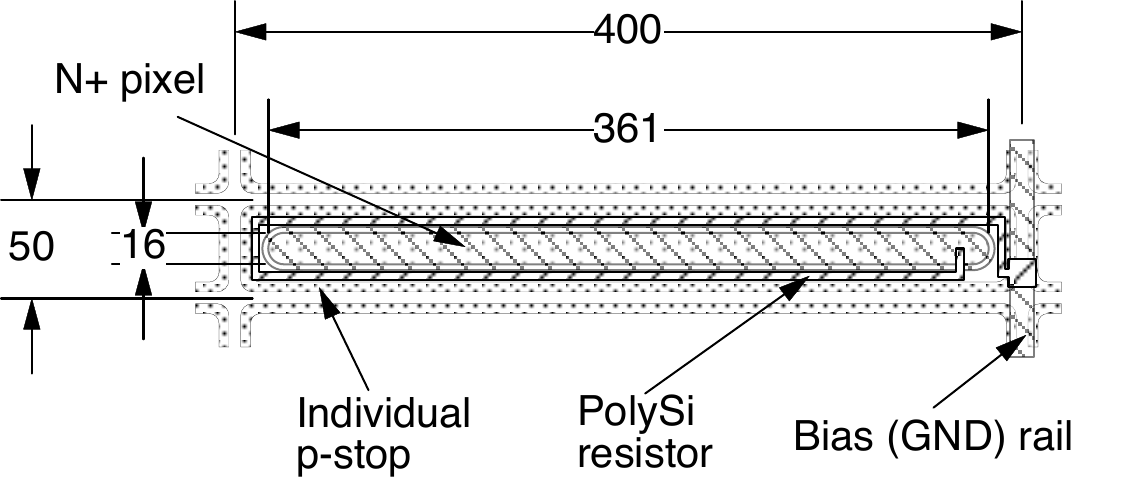} 
\end{center}
\caption{\label{fig:KEK-samples} Pixel cell design details for KEK1 (top) and KEK2 (bottom) samples.}
\end{figure}

Two modules with different sensor n-in-p layouts were subject to beam tests; 
 one with a polysilicon bias resistor and a common p-stop isolation (KEK1), and the other with a polysilicon  bias resistor and an individual p-stop isolation (KEK2). Figure~\ref{fig:KEK-samples} shows a sketch of the pixel cell design for these two samples. 
The sensors came from Float Zone (FZ) wafers with $<$100$>$ crystal orientation; the wafer thickness was 320~$\mu$m.
 The measured wafer resistivity was approximately 6~k$\Omega$cm.  

\noindent A metal bias rail runs along each pixel double-column; there is no bias rail implant underneath it~\cite{HPKsens}. A parylene coating has been applied to the whole body of the pixel modules, after mounted on and connected to the single-chip test card (SCC) with wire-bonding.

These modules were beam-tested before any irradiation in 2010, irradiated afterward and tested in the beamtests in 2011.
 The results before irradiation are reported in this paper; those after irradiation are being analyzed and they will be presented
 in a different communication. 
 The depletion voltage before irradiation was about 180~V.

\subsubsection{Beam test results}

The HPK samples characterization has been carried out by measuring the cluster size, collected charge, charge sharing and spatial resolution as a function of the bias voltage. 

\begin{figure}[!htb]
\begin{center}
\includegraphics[width=0.95\textwidth]{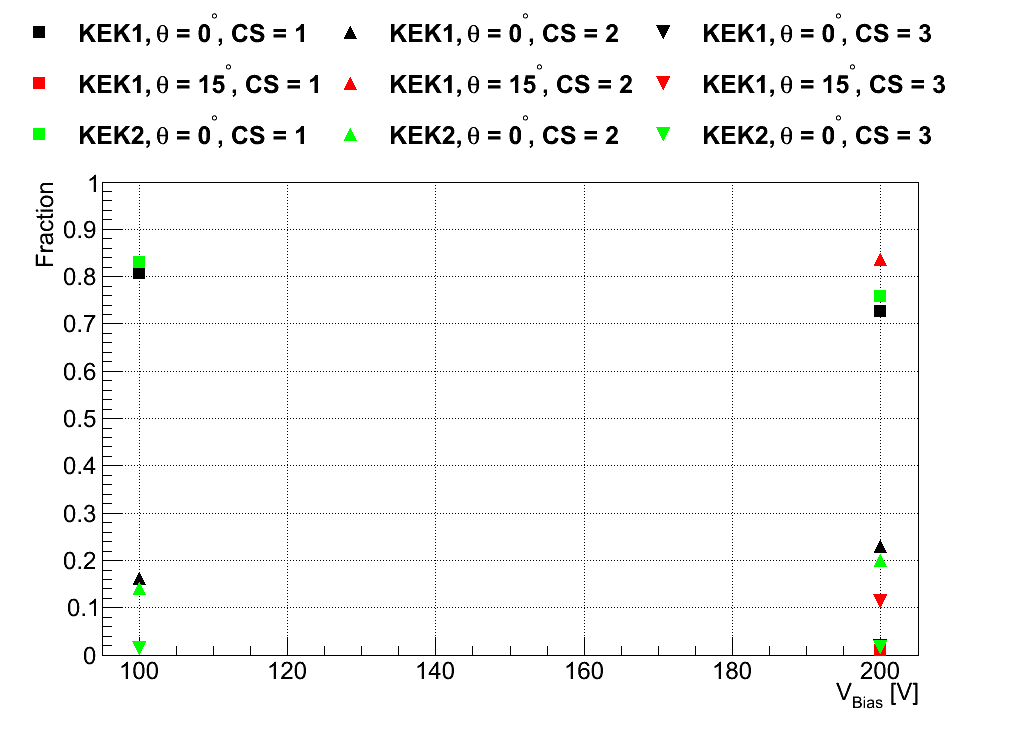}
\caption{\label{fig:KEK-clusterSize}Fraction of cluster sizes as function of the bias voltage for HPK samples; particles were impinging at diffent angles too.}
\end{center}
\end{figure}

\paragraph{Cluster Size}
The KEK1 and KEK2 samples were biased at 100\,V and 200\,V. The two samples perform in a similar way in terms of cluster size for particles at normal track incidence. As shown in Table~\ref{tab:HPK-matchedCSOctober} (see also Figure~\ref{fig:KEK-clusterSize}) more than 80\% of the clusters have just one hit pixel at 100\,V bias voltage. As the bias voltage increases, the fraction of 2-pixel clusters also increases, as more charge is collected. Therefore the charge fluctuations are small compared to the threshold, which reduces the probability of a pixel collecting a signal below threshold. 

\begin{table}[!htb]
\begin{center}
\begin{tabular}{l|c|c|c|c|c}
sample &  $V_{bias}$ [V] & CS=1 [\%] & CS=2 [\%] & CS=3 [\%] & \specialcell[t]{Charge sharing \\probability [\%]}  \\
\hline
\hline
KEK1 & 100 & 81 & 16 & 1 & 18\\
KEK1 & 200 & 73 & 23 & 2 & 26 \\
\hline
KEK2 & 100 & 83 & 14 & 1 & 16\\
KEK2 & 200 & 76 & 20 & 2 & 22 \\
 \end{tabular}
\end{center}
\caption{\label{tab:HPK-matchedCSOctober}Cluster composition for HPK detectors for different bias voltages; charge sharing probability 
 is reported in the last column. Clusters were matched to a track.}
\end{table}

\paragraph{Collected charge}

\begin{figure}[!htb]
\begin{center}
\includegraphics[width=0.85\textwidth]{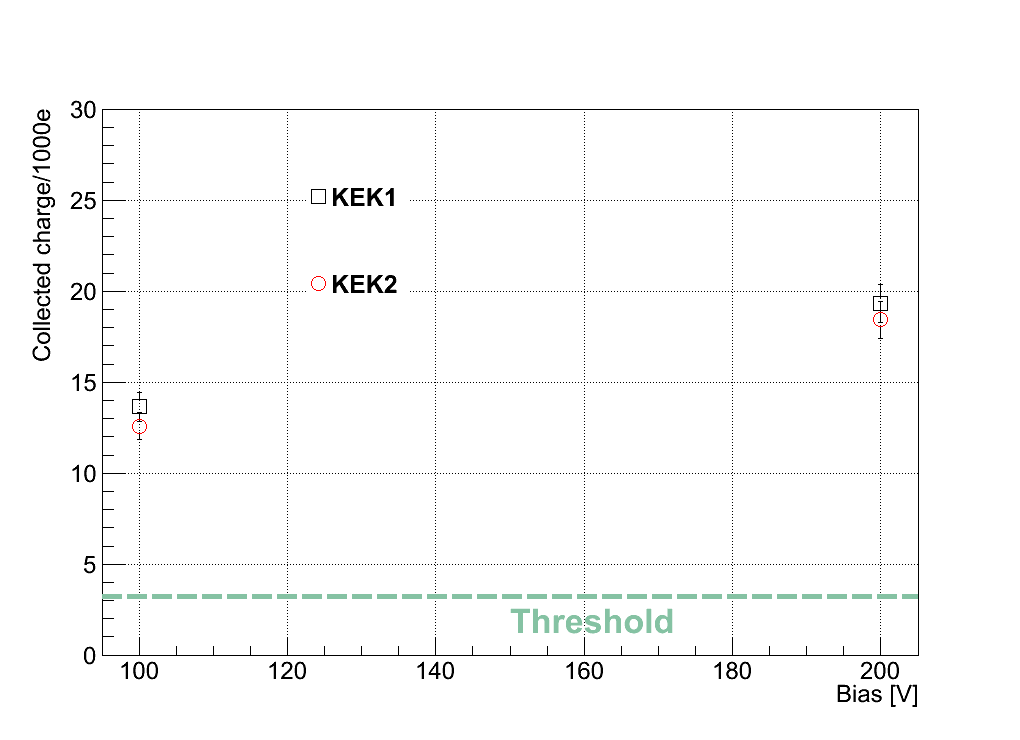}
\caption{\label{fig:KEK-MPV}Collected charge as a function of the bias voltage for HPK samples; particles were impinging at diffent angles too.  A threshold of 3200\,e is indicated.}
\end{center}
\end{figure}

In Figure~\ref{fig:KEK-MPV} the collected charge per cluster is shown as a function of bias voltage for the KEK sensors. The charge collection improves with bias voltage, but already at 100\,V the signal is more than 4 times the threshold. At 200\,V the expected full charge is collected.

\begin{figure}[!htb]
\begin{center}
\includegraphics[width=1\textwidth]{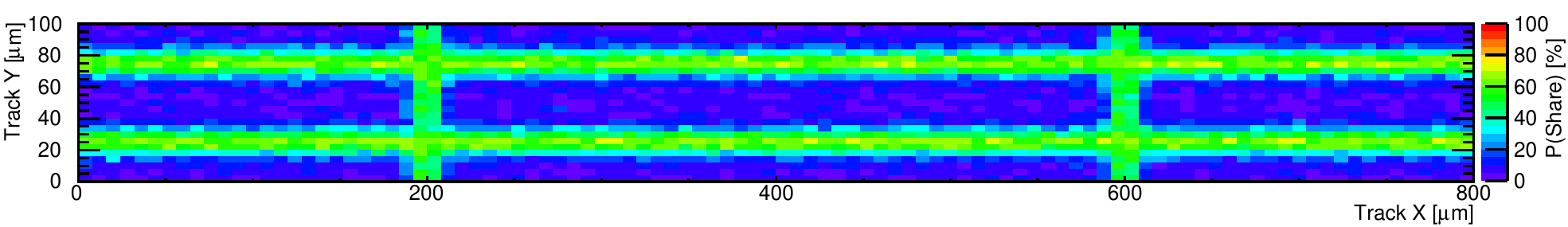}

\includegraphics[width=1\textwidth]{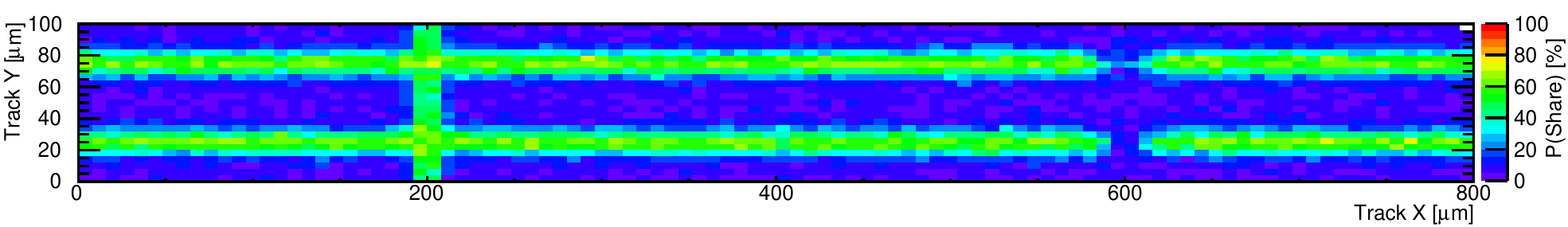}
\caption{\label{fig:KEK-qshare2D}Charge sharing map for KEK1 (top) and KEK2 (bottom) at V$_{bias}$ of 200\,V.}
\end{center}
\end{figure}

\paragraph{Charge sharing}
Figure~\ref{fig:KEK-qshare2D} shows the charge sharing map for KEK1. At normal incidence the fraction of charge sharing is more than 25\,\% for a sensor biased at 200\,V. Results for KEK2 show that the charge sharing is less effective (22\,\%): this can be related to the different layout between the two sensors. In the bottom figure the combined  effect of the bias metal rail and the individual p-stop is visible. 
 Results are summarized in Table~\ref{tab:HPK-matchedCSOctober}.

\begin{figure}[!htb]
\begin{center}
\includegraphics[width=0.45\textwidth]{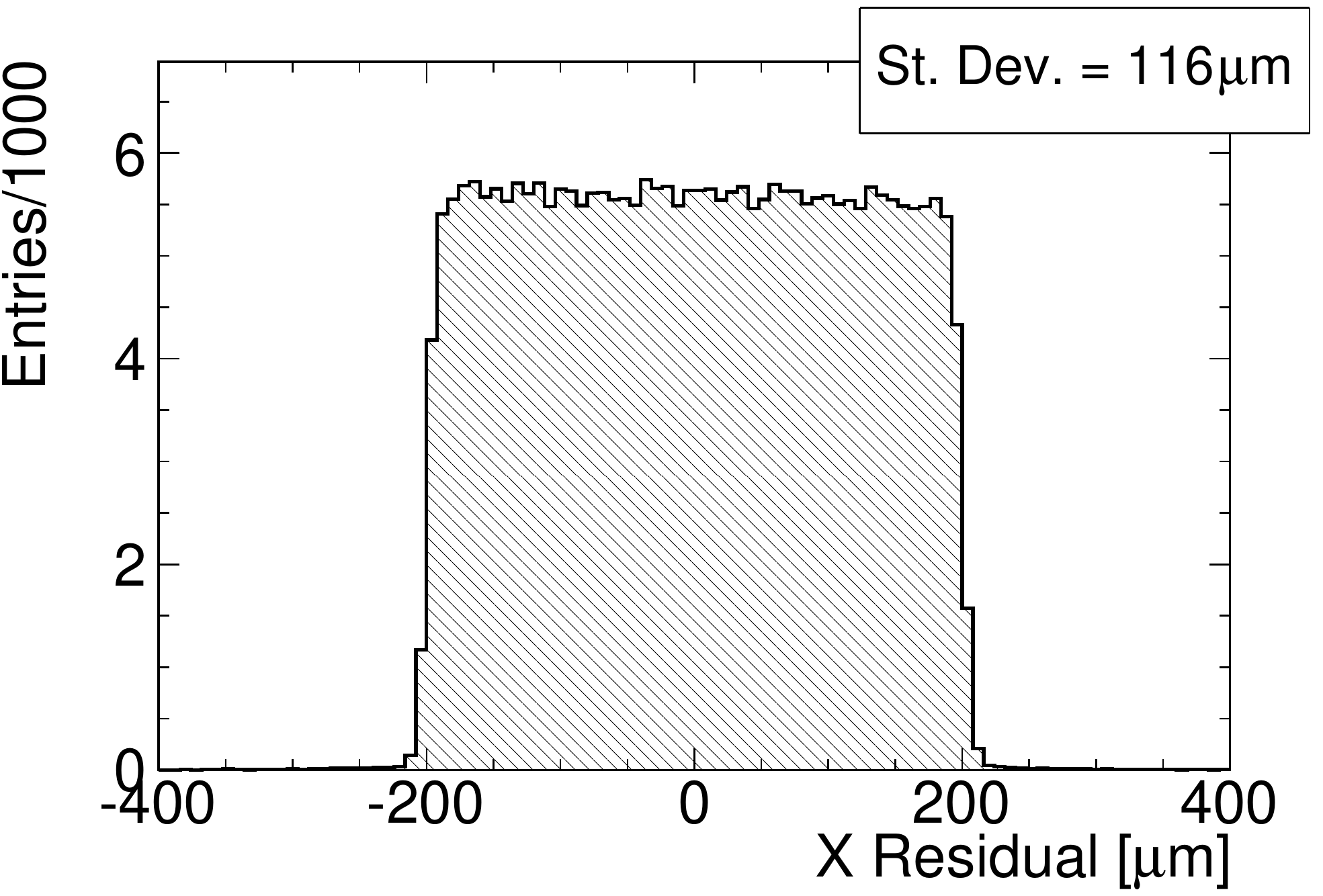} 
\includegraphics[width=0.45\textwidth]{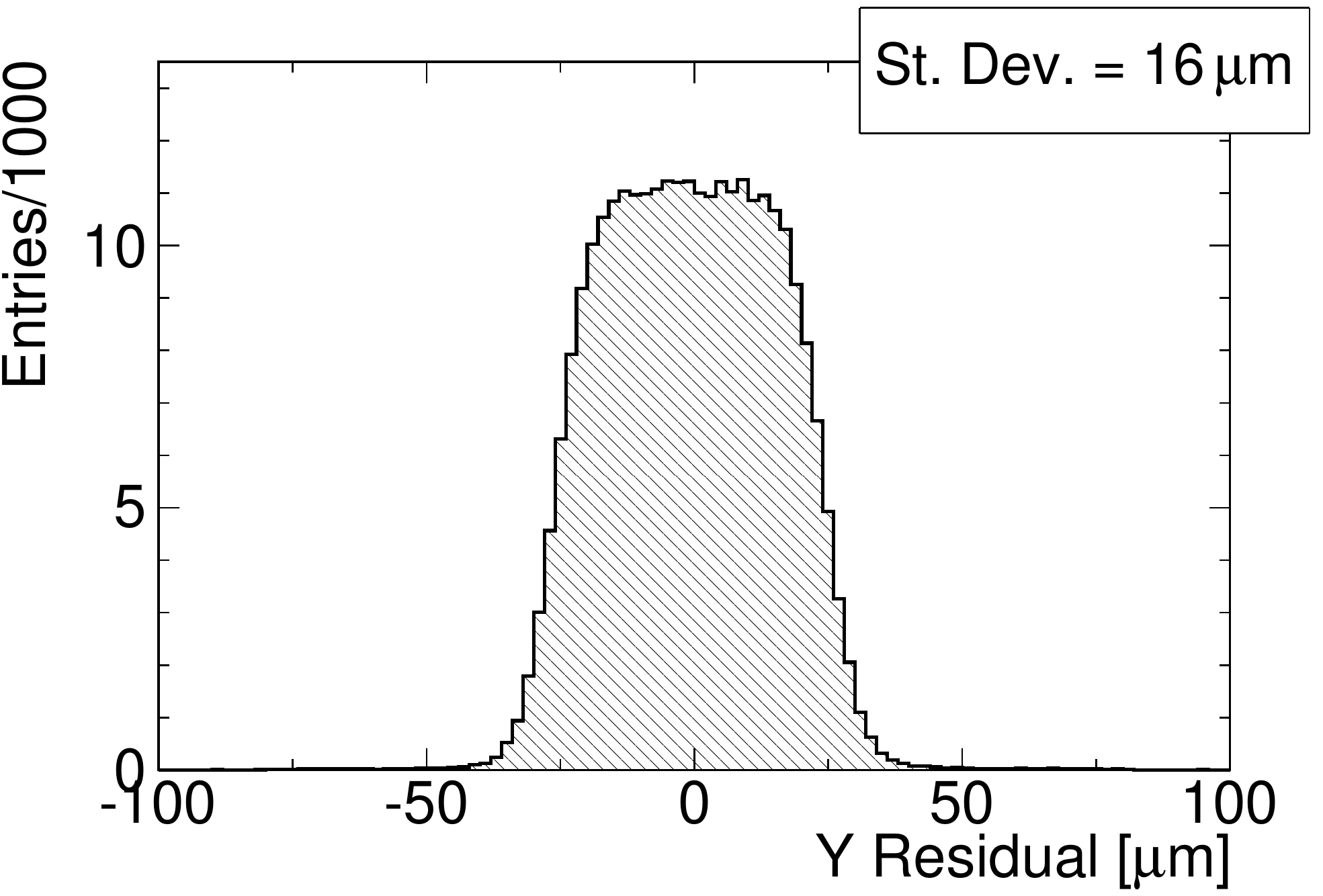} \end{center}
\caption{\label{fig:KEK1_200_resid}Cluster position  residual distribution for non-irradiated sample KEK1 at V$_{bias}$ of 200\,V at normal incidence. 
Left: long pixel projection; right: short pixel projection.}
\end{figure}

\paragraph{Residuals}
Figure~\ref{fig:KEK1_200_resid} shows the residual distribution for all clusters at normal track incidence. The spatial resolution is about 16\,$\mu$m along the short pixel direction, which is comparable with the digital resolution of pitch/$\sqrt{12}$; the same is true for the long  pixel direction (RMS about 116\,$\mu$m). No difference is noticeable between KEK1 and KEK2 sensors. 

\begin{figure}[!htb]
\begin{center}
\includegraphics[width=0.45\textwidth]{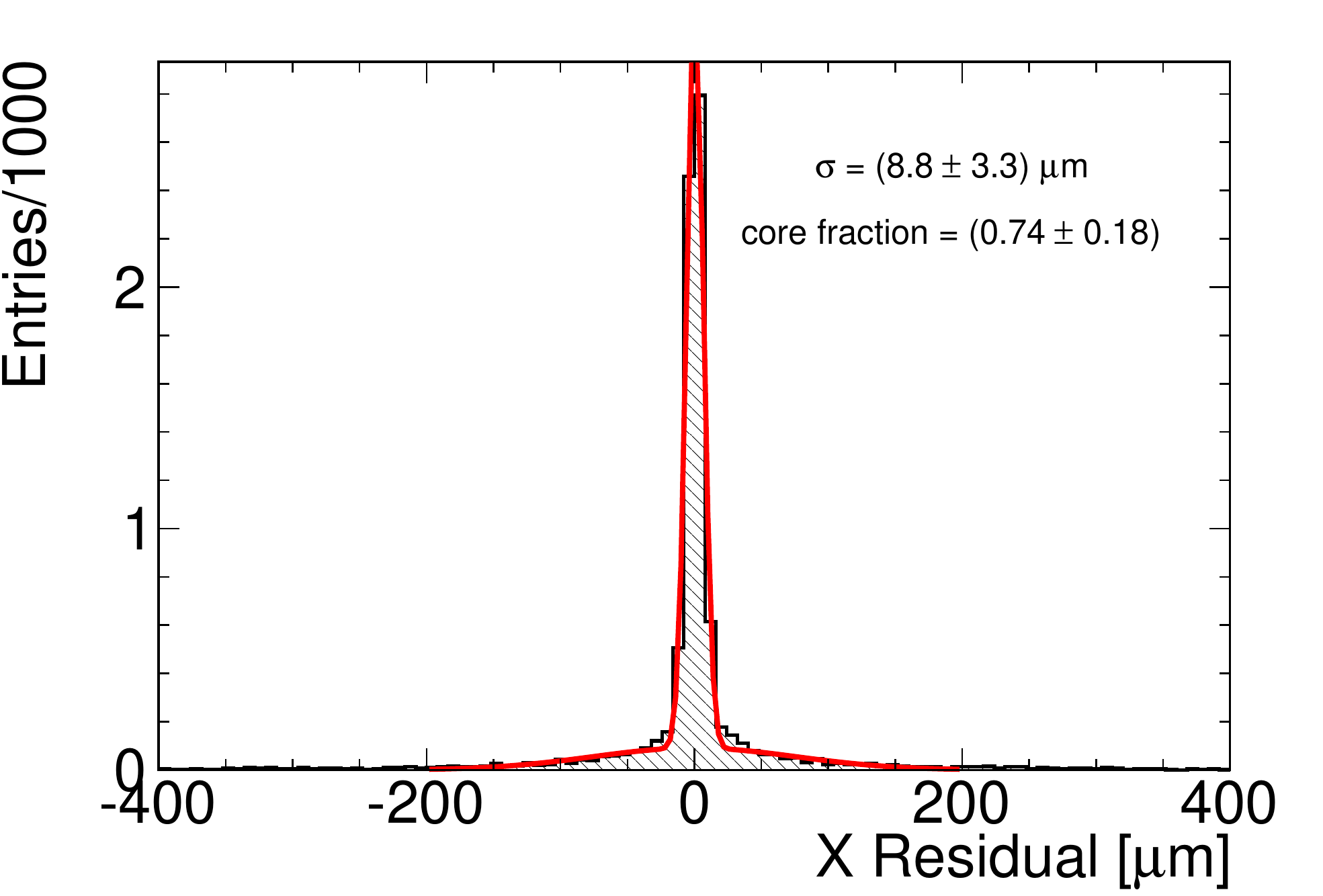} 
\includegraphics[width=0.45\textwidth]{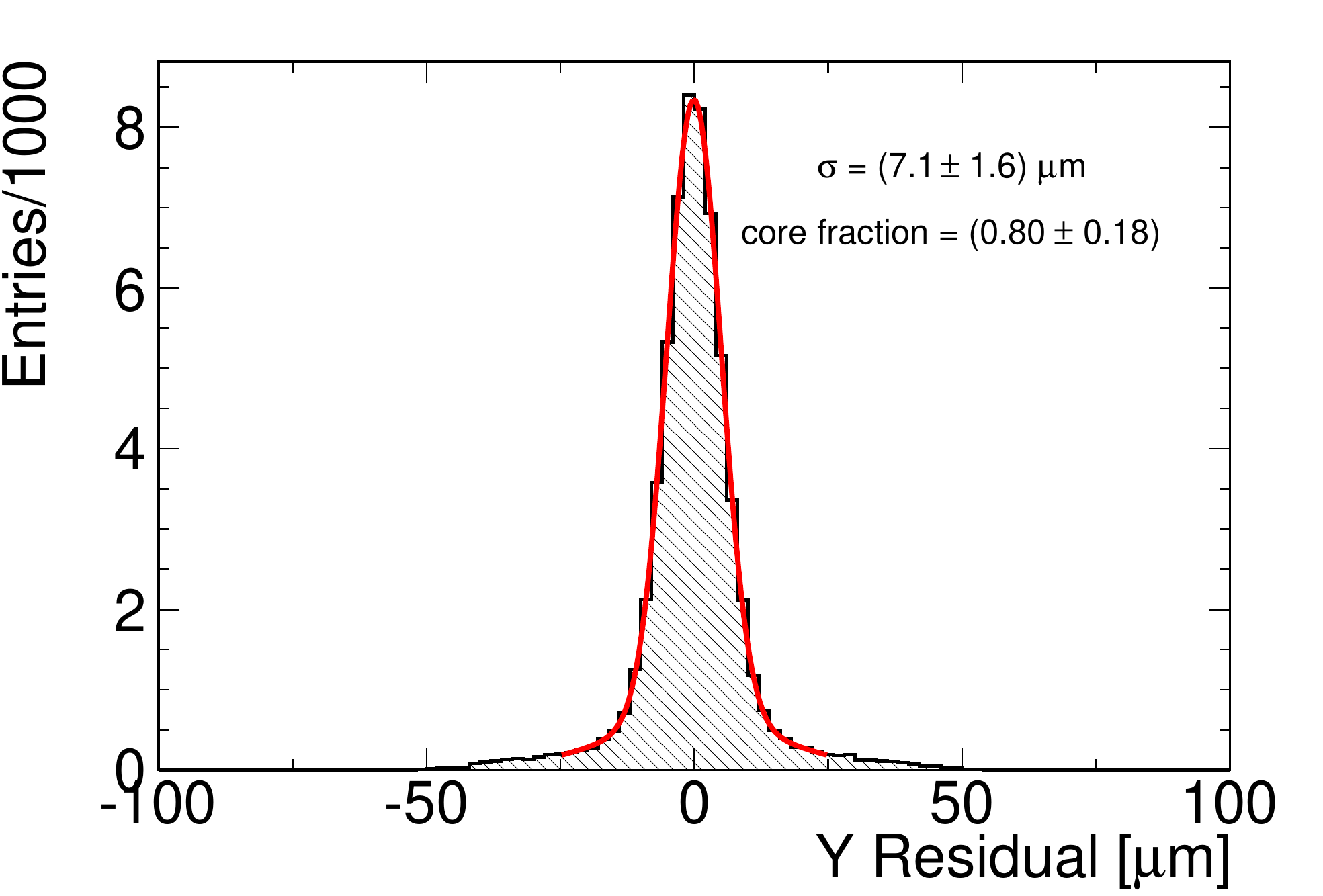} \end{center}
\caption{\label{fig:KEK1_200_resid2}Residual distribution for non-irradiated KEK1 biased at 200 Volts clusters with 2 pixels.  Left: long pixel projection; right: short pixel projection.}
\end{figure}

Figure~\ref{fig:KEK1_200_resid2} shows the residual distribution for two-hit clusters. For these, the spatial resolution is found to be around 7\,$\mu$m in the short pixel direction and around 9\,$\mu$m for the long one (see also Table~\ref{tab:KEK-residual}). The spatial resolution when the cluster contains two hits is larger than the telescope pointing-resolution and gives an estimate  of the charge sharing region between neighbouring pixels.

\begin{table}[!htb]
\begin{center}
\begin{tabular}{c|c|c|c|c|c}
Sample & $V_{bias}$ [V] & pitch ($\mu$m)& RMS ($\mu$m) & core $\sigma$ ($\mu$m) & core fraction [\%] \\
\hline
\hline
\hline
KEK1 & 100 &400& 116.00 $\pm$ 0.14 & 9.00 $\pm$ 0.13 & 68 \\
KEK2 & 100 &400& 115.00 $\pm$ 0.14 & 9.40 $\pm$ 0.20  & 53 \\
\hline
KEK1 & 200 &400& 116.10 $\pm$ 0.20 & 9.0  $\pm$ 3    & 74 \\
KEK2 & 200 &400& 115.00 $\pm$ 0.15 & 9.31 $\pm$ 0.18 & 60 \\
\hline
\hline
KEK1 & 100 & 50& 16.000 $\pm$ 0.020 & 7.14 $\pm$ 0.06 & 78 \\
KEK2 & 100 & 50& 15.800 $\pm$ 0.020 & 6.88 $\pm$ 0.05 & 78 \\
\hline
KEK1 & 200 & 50& 15.90  $\pm$ 0.05  & 7.1  $\pm$ 1.6  & 80 \\ 
KEK2 & 200 & 50& 16.000 $\pm$ 0.021 & 6.81 $\pm$ 0.05 & 79 
\end{tabular}
\end{center}
\caption{\label{tab:KEK-residual}Summary of residual results for KEK1 and KEK2  samples. Core $\sigma$ and fraction are evaluated for 2-pixels clusters only.}
\end{table}


\section{Radiation hardness of \ninn\ sensors}
\label{sec:n-in-n}

The n-in-n sensor technology used in the current ATLAS Pixel detector have been tested to fluences up to 
$1.1\times10^{15}$\,n$_{\rm eq}$/cm$^2$~\cite{Attilio}. To evaluate the usability of n-type bulk material for IBL and 
future detector upgrades, sensors have been irradiated with fluences as high as $2\times10^{16}$\,n$_{\rm eq}$/cm$^2$ using both reactor neutrons (Jo$\check{\mathrm{z}}$ef-Stefan Institute, Ljubljana)~\cite{Ljubirrad} and protons 
of 25\,MeV (Karlsruhe Institute of Technology - KIT)~\cite{KIT} or 24\,GeV (CERN)~\cite{IRRAD1}.

Most of the sensors follow the ATLAS Pixel detector sensor design with 16 guard rings and a thickness of 250\,$\mu$m. DO6 is a special sensor with only 11 guard rings overlapping the pixel region, designed to study possibilities to reduce the inactive area at the edge of the sensor (see section \ref{sec:slim}) and produced on 285$\mu$m thick bulk material.
The n-type sensors were produced at CiS, from Diffusion Oxygenized Float Zone (DOFZ), $<$111$>$ oriented wafers;
 the wafer resistivity was in the range between 
2 and 5~k$\Omega$. The depletion voltage was in the range 
between 40 and 100~V for 250~$\mu$m thick sensors and between 50 and 140~V for 285~$\mu$m thick sensors.
 The inter-pixel isolation is achieved by means of a ``moderated'' p-spray implantation~\cite{pixel-electronics-paper}.

A total of 5 irradiated n-in-n pixel sensors were tested. Table~\ref{tab:n-in-n} summarizes the fluences to which the sensors were irradiated.
 See also~\cite{Cooling}.

\begin{table}[!htb]
\begin{center}
\begin{tabular}{l|c|c|l}
name & thickness ($\mu$m)  & fluence ($10^{15}$\,n$_{\rm eq}$/cm$^2$) & irradiation type\\
\hline \hline
DO6 & 285 & 0 & -- \\
DO7 & 250 & 1 & protons (KIT)  \\
DO8 & 250 & 1 & reactor neutrons  \\
DO9 & 250 & 5 & reactor neutrons \\ 
DO10 & 250 & 20 & reactor neutrons \\ 
\hline
\end{tabular}
\end{center}
\caption{\label{tab:n-in-n}Summary of irradiated n-in-n samples in the testbeams. KIT stands for 25\,MeV energy proton irradiation.
}
\end{table}

\subsection{Results}
\label{sec:n-in-n:Results}

Measurements on n-in-n samples have been carried out at temperatures well 
below 0$^{\circ}$~C to reduce the large leakage current from irradiated 
sensors. As an example, we measured a leakage current of 24~${\rm \mu A}$ 
 (10~${\rm \mu A}$) for DO10 (DO9), at a bias voltage of 1200~V and at 
 -47$^{\circ}$~C.

\begin{figure}[!tb]
 \begin{center}
  \includegraphics[width=\textwidth]{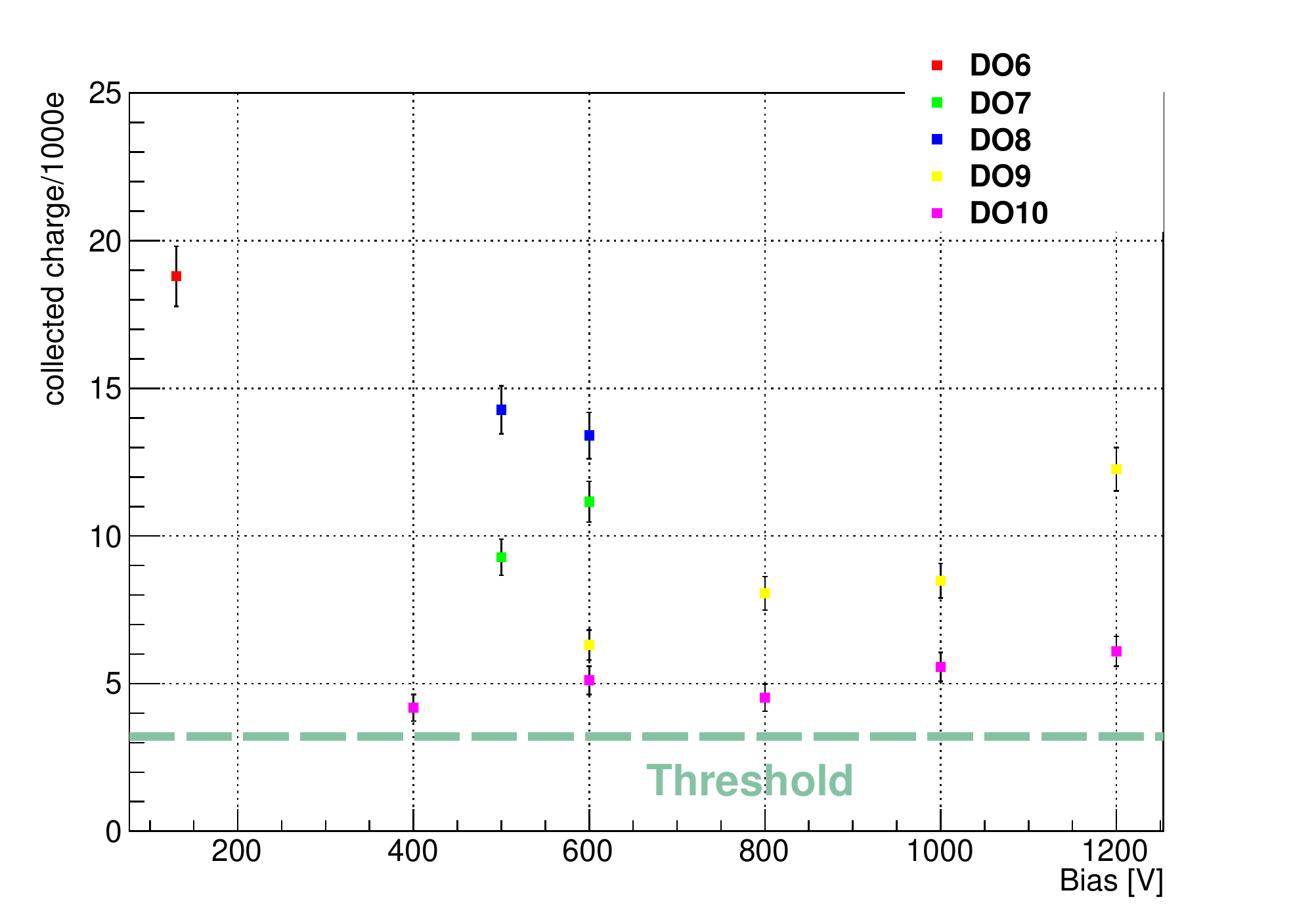}
 \end{center}
 \caption{\label{fig:n-in-n:CCE-Overview}Collected charge as a function of bias voltage for n-in-n samples irradiated to different fluences
 (see details in the text).
 A threshold of 3200~e is indicated.}
\end{figure}

\paragraph {Charge collection}
One of the main effects of irradiation is the increased trapping, which leads to a reduced signal amplitude. As the trapping probability depends on the charge carrier velocity, the collected charge was measured as a function of the bias voltage. Figure~\ref{fig:n-in-n:CCE-Overview} shows the results for all irradiated n-in-n samples in the two beam test periods; see also Table~\ref{tab:n-in-n}. A systematic error on the collected charge of 400\,e is assumed, due to the finite charge resolution of the ToT mechanism;  a 5\% systematic uncertainty is also taken into account, due to non-uniformity in the injection capacitances.\\
After $5\times 10^{15}$\,n$_{\rm eq}$/cm$^2$, the collected charge still exceeds 10~ke at a bias voltage of 1000\,V. Even if the collected charge is shared equally between two neighboring pixels, this charge is sufficient to detect the hit with FE-I3.

\begin{figure}[!htb]
 \begin{center}
  \includegraphics[width=1\textwidth]{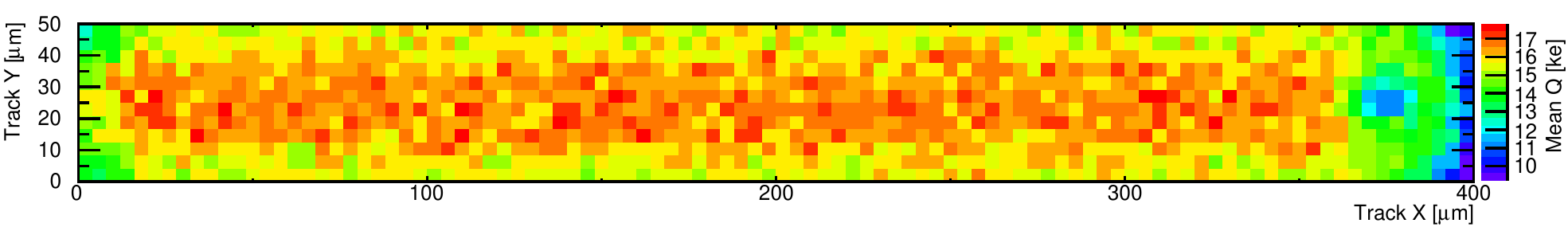}
  \includegraphics[width=1\textwidth]{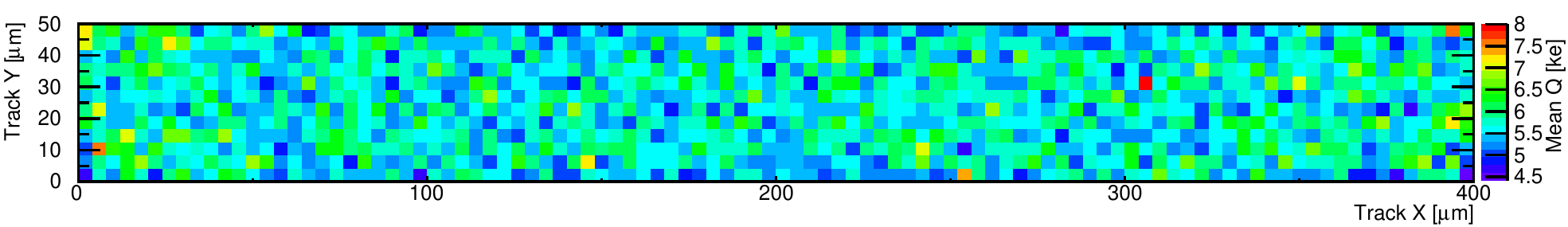}
 \end{center}
 \caption{Charge collection within a pixel. Top: DO9 at V$_{bias}$=1200\,V. Bottom: DO10 at V$_{bias}$=1000\,V.\label{fig:n-in-n:Oct_P5_D11_qeff}}
\end{figure}

Figure \ref{fig:n-in-n:Oct_P5_D11_qeff} top, shows that charge is predominantly
 lost in the region of the punch-through bias grid system.

 At very high fluences ($2\times 10^{16}$\,n$_{\rm eq}$/cm$^2$, DO10 sample)
 it is no longer possible to say which region is less efficient than the others, using the charge collection method 
 (Figure \ref{fig:n-in-n:Oct_P5_D11_qeff}, bottom).

\begin{figure}[!htb]
 \begin{center}
 \hspace{-.28cm}
  \includegraphics[width=0.887\textwidth]{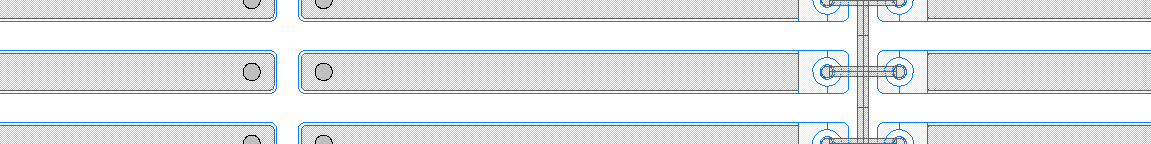}
  \includegraphics[width=\textwidth]{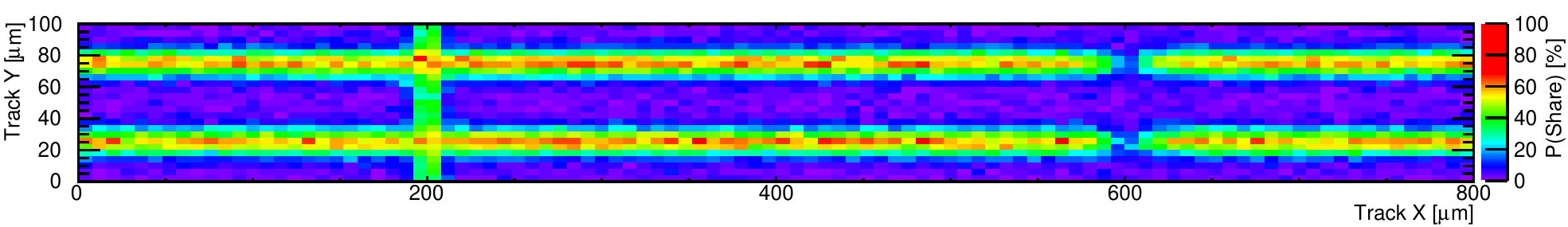}
 \end{center}
 \caption{Top: Design of the sample of the region shown in the plot below. Bottom: Charge sharing probability for DO9 at V$_{bias}$=1200V. Note the reduced charge sharing in the bias grid region on the right-hand side of the central pixel. \label{fig:n-in-n:DO9_qshare}}
\end{figure}

\paragraph {Charge sharing}
Figure \ref{fig:n-in-n:DO9_qshare} shows the charge sharing probability for DO9 at a bias voltage of 1200\,V.
Reduced charge sharing probability is visible in the region of the bias dot
 and the bias grid network.\footnote{The bias grid network is an aluminum trace
 arranged on top of the intermediate pixel region connecting all bias dots.}
 Less charge is deposited here, so there is a higher probability for the second pixel in a two-pixel cluster to be below threshold.
As only the bias trace makes the difference between both pixel sides, it might cause the lower charge sharing probability. Furthermore, one can see that the region of the bias dot is not affected.

\begin{figure}[!htb]
 \begin{center}
  \includegraphics[width=\textwidth]{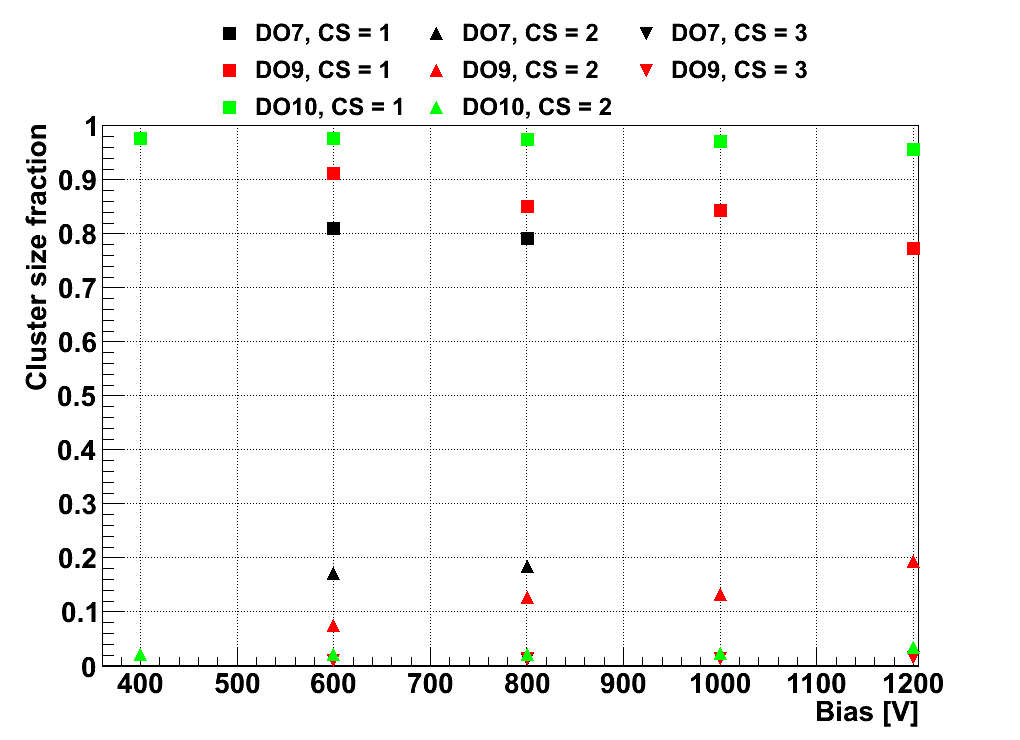}
 \end{center}
 \caption{\label{fig:n-in-n:CSfraction}Fractions of 1-, 2-, and 3-hit clusters as function of bias voltage for irradiated n-in-n samples; see text for details.
 Error bars are too small to be visible.}
\end{figure}

 While for DO9 a clear increase in charge sharing probability towards the edges of the pixel is visible, at higher fluence the collected charge becomes too small for any significant charge sharing to be observable. This can also be seen in the fractions of clusters with one, two, and more pixels.
Figure \ref{fig:n-in-n:CSfraction} shows the fractions of one-pixel, two-pixel, and larger clusters as a function of the bias voltage.
 It is evident, that with increasing bias voltage the cluster size increases, due to the reduced trapping. At a given voltage the fraction of 1-pixel clusters increases with fluence, as more charge is lost due to trapping. \\
For samples irradiated up to $5\times 10^{15}$\,n$_{\rm eq}$/cm$^2$ the cluster size increases slightly with bias voltage, while at $2\times 10^{16}$\,n$_{\rm eq}$/cm$^2$ the fraction of clusters with two or more hit pixels is very small and stays nearly constant over the accessible voltage range.

\begin{figure}[!hbt]
 \begin{center}
  \includegraphics[width=.49\textwidth]{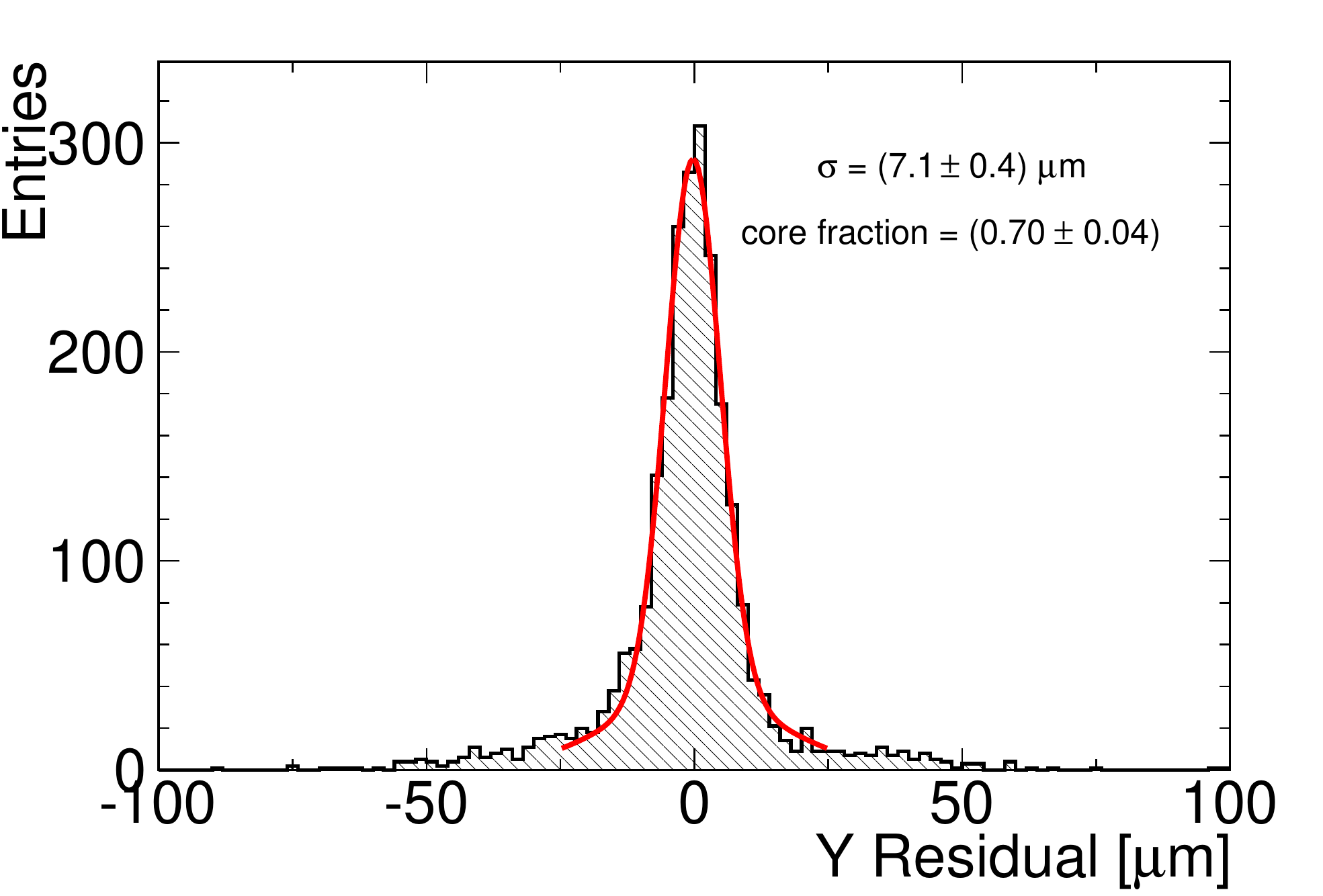}
  \includegraphics[width=.49\textwidth]{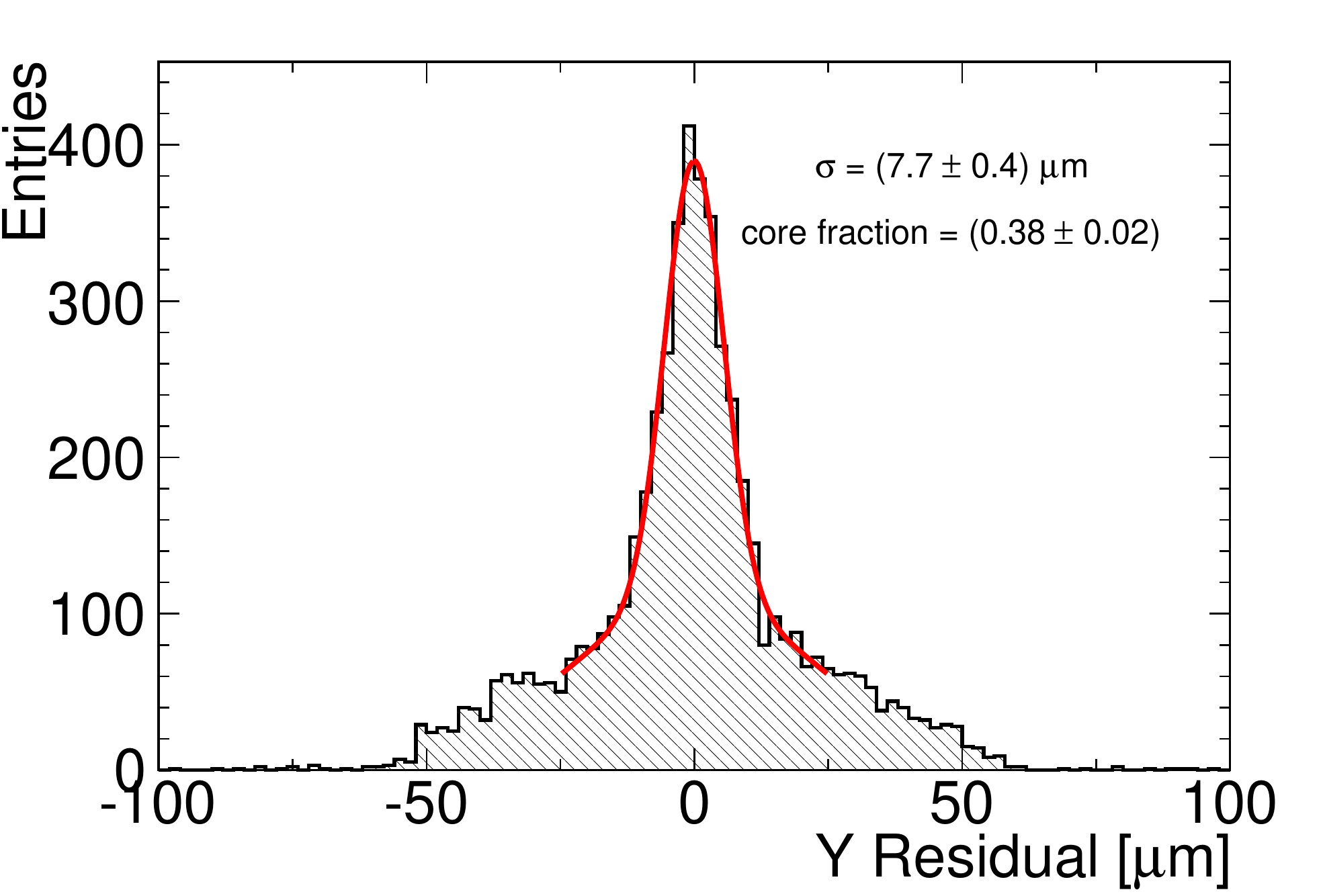}
 \end{center}
 \caption{Residual distributions for 2-pixel clusters only. Shown are distributions samples irradiated to $5\times 10^{15}$\,n$_{\rm eq}$/cm$^2$ (left: DO9, bias voltage 1000\,V) and $2\times 10^{16}$\,n$_{\rm eq}$/cm$^2$ (right, DO10, bias voltage 1200\,V), respectively. \label{fig:n-in-n:CS2-residuals}}
\end{figure}

Plotting the residual distribution for two-pixel clusters only allows the width of the charge sharing region between pixels to be determined. Figure \ref{fig:n-in-n:CS2-residuals} shows the distributions for DO9 ($5\times 10^{15}$\,n$_{\rm eq}$/cm$^2$) and DO10 ($2\times 10^{16}$\,n$_{\rm eq}$/cm$^2$). After correcting for the telescope resolution, the widths of the charge sharing regions are 7.1\,$\mu$m and 7.7\,$\mu$m. These values correspond very well with the width found for an unirradiated sample of 6.4\,$\mu$m. This indicates that the lateral diffusion of the charge cloud does not change significantly with irradiation.

\begin{figure}[!htb]
 \begin{center}
  \includegraphics[width=.49\textwidth]{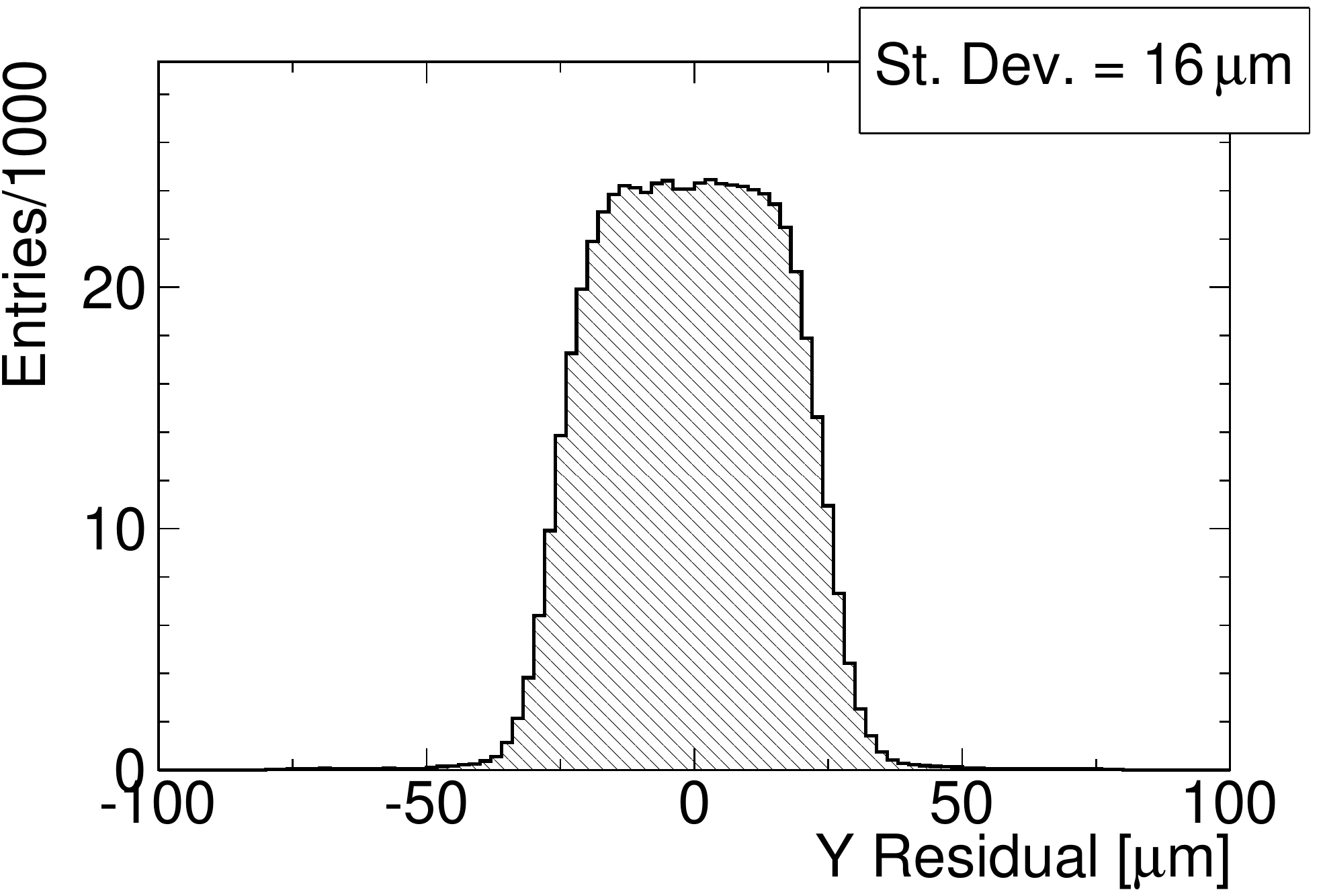}
  \includegraphics[width=.49\textwidth]{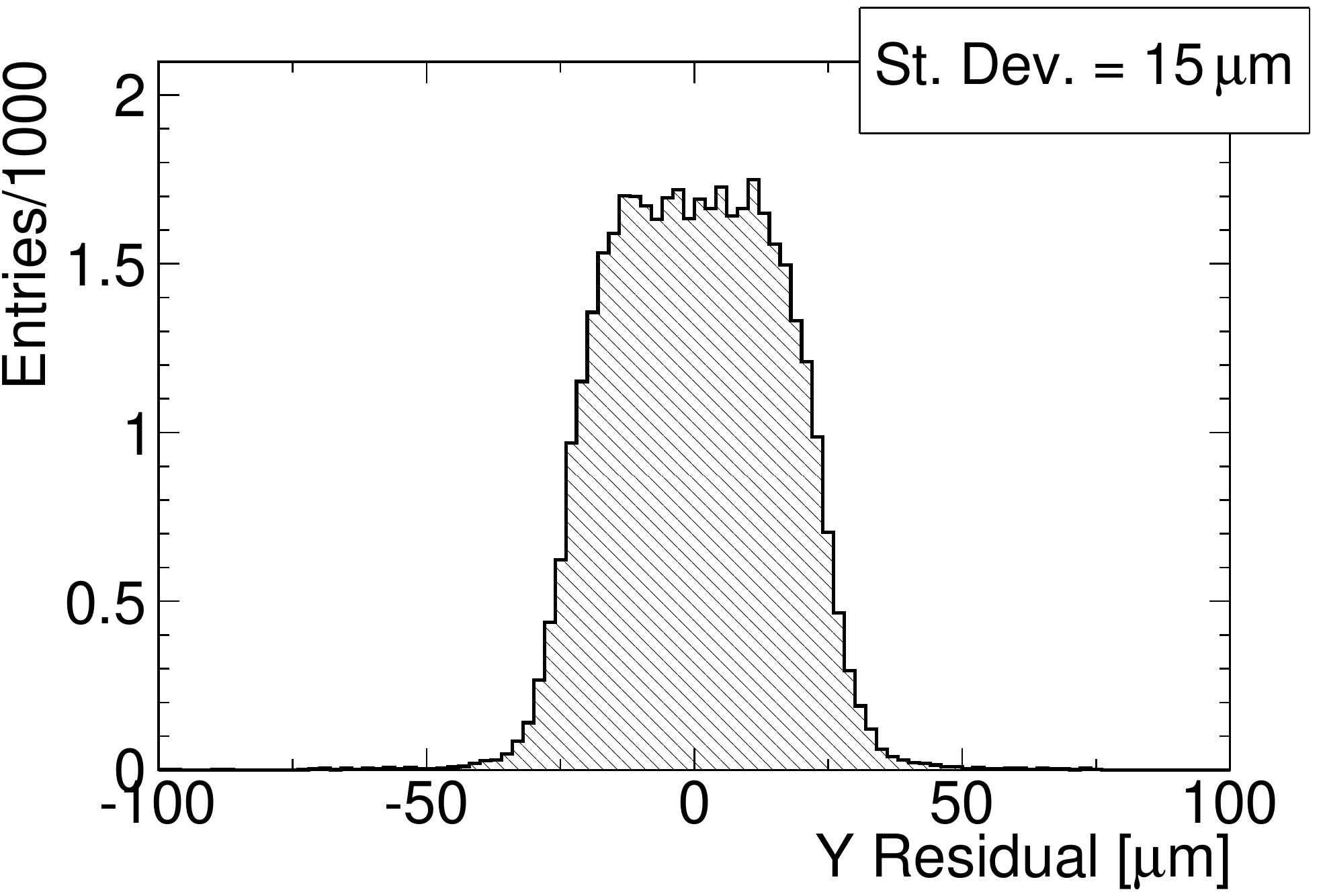}
 \end{center}
 \caption{Residual distributions in the short pixel direction for an unirradiated sample (DO6, left) and a sample irradiated to $2\times 10^{16}$\,n$_{\rm eq}$/cm$^2$ operated at a bias voltage of 1000\,V (DO10, right). No deterioration of the spatial distribution with irradiation is visible. \label{fig:n-in-n:residuals}}
\end{figure}

\paragraph {Residuals}
Figure \ref{fig:n-in-n:residuals} shows the residual distributions in the 50\,$\mu$m pixel direction for the unirradiated sample (DO6) and the sample irradiated to 
\mbox{$2\times10^{16}\,{\rm n_{eq}}/{\rm cm}^2$}, respectively. The widths of the distributions are 16\,$\mu$m and 15.4\,$\mu$m, comparable with the expected digital resolution of 14.4\,$\mu$m. Thus, no influence of radiation damage on the spatial resolution can be observed.

\section{Slim Edge}
\label{sec:slim}

For slim edge studies the outermost pixels of a sample are of special interest. Therefore, the samples were mounted such that the edge of the sensor was well within the trigger acceptance window.\\ Special analysis classes were written to investigate the characteristics of the edge pixels. The basic principle is the same as for the charge collection analysis but instead of overlaying all pixels onto one single pixel, only pixels at the sensor edge are used and the special geometry is conserved in the overlay process.

\begin{figure}[!htb]
 \begin{center}
  \includegraphics[width=0.6\textwidth]{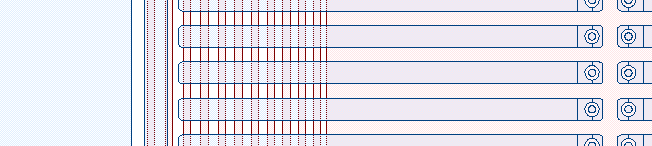}
  \includegraphics[width=0.6\textwidth]{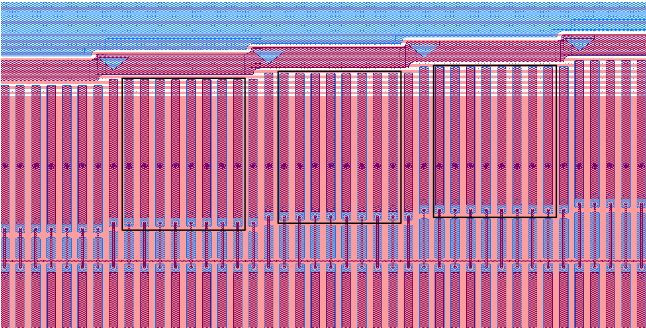}
 \end{center}
 \caption{Test structures for slim edge studies. Top: DO6, the active pixel region overlaps the guard rings by 210\,$\mu$m. Bottom: DO3, groups of 10 pixels shifted towards the edge of the sensor in steps of 25\,$\mu$m. \label{fig:slim}}
\end{figure}

For the IBL sensors the width of the inactive region at the edge of each sensor tile has to be reduced significantly with respect to the approximately 1\,mm wide region on each side of the current ATLAS Pixel detector sensors. One approach is to shift the guard-rings on the p$^+$-side inwards. Two specially designed DUTs were tested to study the impact of 
an overlap between the active pixel region with the guard ring region, where the electric field in the sensor is inhomogeneous.\\
In the DO6 sample, the overlap between active pixel region and guard ring region is 210\,$\mu$m, with the number of guard rings reduced to 11. In the DO3 sample groups of 10 pixels are shifted towards the edge of the sensor in steps of 25\,$\mu$m, increasing the area in which the pixels overlap with the guard-rings (see Figure~\ref{fig:slim}).
 
\begin{figure}[!htb]
 \begin{center}
  \includegraphics[width=\textwidth]{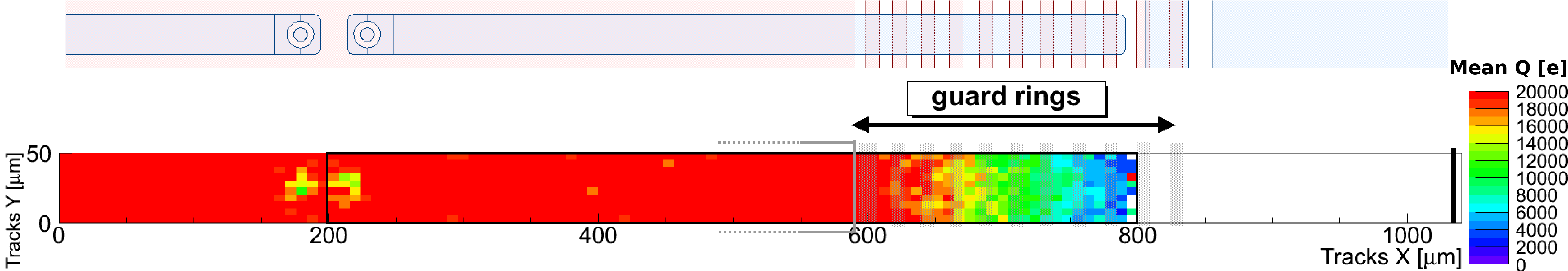}
 \end{center}
 \caption{Charge collection in the pixels shifted underneath the guard rings.\label{fig:CCE-DO6}}
\end{figure}

The test structures were mounted such that the edge of the sensor was well in the center of the trigger window, allowing to study charge collection in the shifted pixels in some detail. Figure \ref{fig:CCE-DO6} shows the collected charge in the overlap region of DO6. With increasing distance from the bias voltage pad the collected charge decreases, due to the inhomogeneously formed depletion zone. 
 It is evident that the collected charge is sufficient to ensure good hit efficiency up to about 200\,$\mu$m from the edge of the bias voltage pad.\\

\begin{figure}[!htb]
 \begin{center}
  \includegraphics[width=\textwidth]{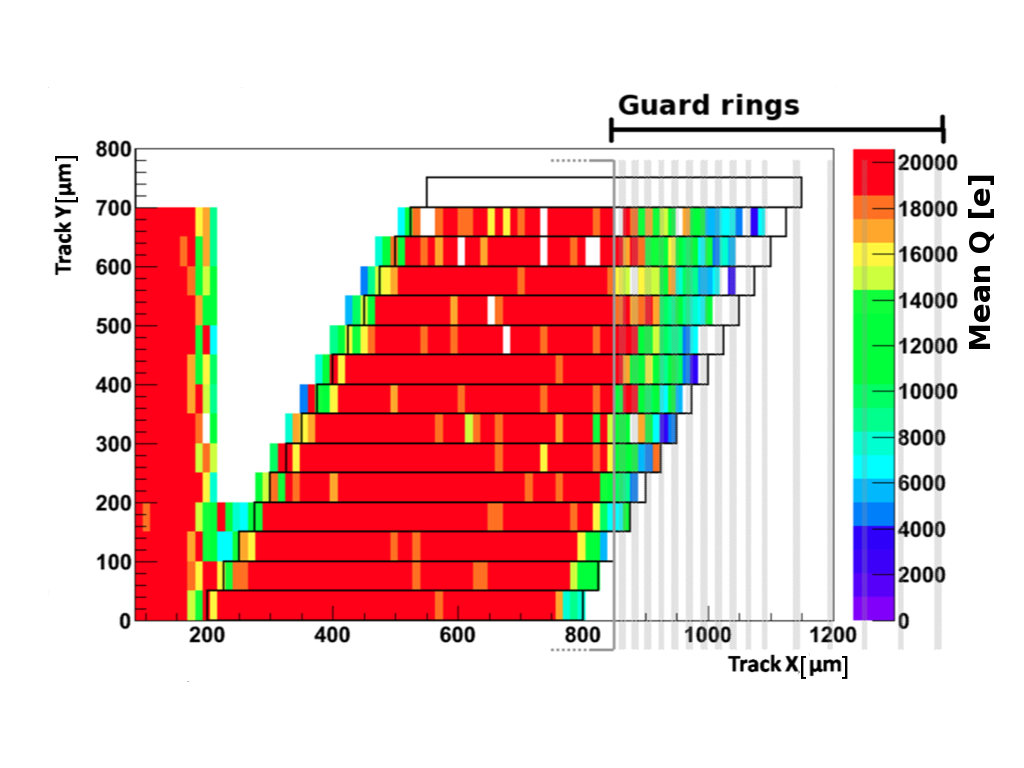}
 \end{center}
 \caption{Charge collection in the pixels shifted underneath the guard rings.\label{fig:CCE-slim}}
\end{figure}

Figure \ref{fig:CCE-slim} shows the collected charge in the overlap region for the DO3 sample. Data from pixels with the same shift with respect to the edge of the bias voltage pad are plotted into one pixel. The drop in collected charge systematically occurs at the same distance from the bias voltage pad, regardless of the shift of the pixel. This indicates, that the loss of collected charge is indeed due to the depletion zone which is expected to be inhomogeneous along the x-axis  (orthogonal to the bias voltage pad) but homogeneous along the y-axis (parallel to the bias voltage pad). 
 Further studies of this kind can be found in~\cite{PSD9Slim}.

\section{Conclusion}
\label{sec:conclusion}

Planar silicon sensors,   have been tested in high energy pion beams at the CERN SPS North Area in 2010 by the ATLAS Planar Pixel Sensors (PPS) collaboration. Different bulk materials, geometries, especially for the guard ring regions, biasing and isolation structures were examined.\\ 
The goals of the measurement program were threefold: to demonstrate the suitability of p-bulk sensors for tracking purposes, to prove the radiation hardness of n-bulk sensors and to realize pixel sensors with reduced inactive edge area.

Pixelated p-bulk sensors produced by different vendors were tested to evaluate their performance, after irradiation too.
 In terms of the collected charge, charge sharing, and spatial resolution the performance of the p-bulk sensors was very good and comparable to that of n-bulk sensors. The issue of the high potential on the pixelated side of the sensor was tested and operation of the sensors was proven to be very stable.

The radiation hardness of n-bulk sensors was tested up to unprecedented fluences, with a maximum of $20\times10^{15}\,{\rm n_{eq}}/{\rm cm}^{2}$.
 At a bias voltage of 1.2\,kV a collected charge of about 6\,ke was observed, corresponding to about one third of the collected charge before irradiation. Despite the rather small collected charge and the reduced charge sharing between pixels, no significant deterioration of the
 spatial resolution was observed.

In order to reduce the inactive area at the edge of n-bulk sensors, several modified sensor layouts were tested. The influence of a reduction of the number of guard rings and an increasing overlap between the active pixel region and the guard ring region on the backside of the sensor were studied. It was found that the charge collection efficiency reduces with increasing distance from the edge of the bias voltage pad due to the inhomogeneously formed depletion zone in the sensor.
However, the collected charge is sufficient for reliable particle detection up to a distance of about 200\,$\mu$m from the bias voltage pad. This was very encouraging for the planar ATLAS IBL candidate design, which was finally designed employing the methods evaluated by the beam test measurements described in this paper.

\newpage

\section*{Acknowledgements}
The authors would like to express their gratitude to V.~Cindro, G.~Kramberger and I.~Mandi\'{c} for their valuable support with irradiations at the TRIGA reactor of the Jo\v{z}ef Stefan Institute, Ljubljana, to A.~Dierlamm for his help at the Irradiation Center, Karlsruhe, and to M.~Glaser for his help at the CERN PS irradiation facility.

The work has been partially performed in the framework of the CERN RD50 Collaboration. This work is supported by the Commission of the European Communities under the 6th Framework Programme ``Structuring the European Research Area'', contract number \protect{RII3-026126}.

We gratefully acknowledge the financial support of the German Federal Ministry of Science and Education (BMBF) within their excellence program, in particular as part of the collaborative research center ``FSP 101-ATLAS, Physics on the TeV-scale at the Large Hadron Collider''.

We acknowledge the support of the Initiative and Networking Fund of the Helmholtz Association, contract \protect{HA-101} (``Physics at the Terascale'').


\end{document}